\documentclass[fleqn,usenatbib]{mnras}

\usepackage[T1]{fontenc}

\DeclareRobustCommand{\VAN}[3]{#2}
\let\VANthebibliography\thebibliography
\def\thebibliography{\DeclareRobustCommand{\VAN}[3]{##3}\VANthebibliography}

\usepackage{graphicx}	
\usepackage{amsmath}	
\usepackage{amssymb}	
\usepackage{bm}	
\usepackage{newtxtext,newtxmath}

\newcommand{\kpc}{\hbox{$\mathrm{kpc}$}}
\newcommand{\kms}{\hbox{$\mathrm{km \; s^{-1}}$}}
\newcommand{\kmskpc}{\hbox{$\mathrm{km \; s^{-1} \; kpc^{-1}}$}}
\newcommand{\OmegaBar}{\hbox{$\Omega_{\mathrm{b}}$}}

\title[Bar resonances in an $N$-body Galactic disc]{Impact of bar resonances in the velocity-space distribution of the solar neighbourhood stars in a self-consistent $N$-body Galactic disc simulation}

\author[T. Asano et al.]{
Tetsuro Asano,$^{1}$\thanks{E-mail: t.asano@astron.s.u-tokyo.ac.jp}
Michiko S. Fujii,$^{1}$
Junichi Baba,$^{2}$
Jeroen B{\' e}dorf,$^{3,5}$
Elena Sellentin$^{3,4}$, and 
\newauthor
Simon Portegies Zwart$^{3}$
\\
$^{1}$Department of Astronomy, Graduate School of Science, 
    The University of Tokyo, 7-3-1 Hongo, Bunkyo-ku, Tokyo, 113-0033, Japan\\
$^{2}$National Astronomical Observatory of Japan, Mitaka-shi, Tokyo 181-8588, Japan\\
$^{3}$Leiden Observatory, Leiden University, NL-2300RA Leiden, The Netherlands\\
$^{4}$Mathematical Institute, Leiden University, NL-2300RA Leiden, The Netherlands\\
$^{5}$Minds.ai, Inc., Santa Cruz, the United States
}

\date{Accepted XXX. Received YYY; in original form ZZZ}

\pubyear{2021}

\begin{document}
\label{firstpage}
\pagerange{\pageref{firstpage}--\pageref{lastpage}}
\maketitle

\begin{abstract}
The velocity-space distribution of the solar neighbourhood stars shows complex substructures. Most of the previous studies use static potentials to investigate their origins. Instead we use a self-consistent $N$-body model of the Milky Way, whose potential is asymmetric and evolves with time. 
In this paper, we quantitatively evaluate the similarities of the velocity-space distributions in the $N$-body model and that of the solar neighbourhood, using Kullback-Leibler divergence (KLD). 
The KLD analysis shows the time evolution and spatial variation of the velocity-space distribution. 
The KLD fluctuates with time, which indicates the velocity-space distribution at a fixed position is not always similar to that of the solar neighbourhood. 
Some positions show velocity-space distributions with small KLDs (high similarities) more frequently than others. One of them locates at $(R,\phi)=(8.2\;\kpc, 30^\circ)$, where $R$ and $\phi$ are the distance from the galactic centre and the angle with respect to the bar's major axis, respectively.
The detection frequency is higher in the inter-arm regions than in the arm regions.
In the velocity maps with small KLDs, we identify the velocity-space substructures, which consist of particles trapped in bar resonances. The bar resonances have significant impact on the stellar velocity-space distribution even though the galactic potential is not static.
\end{abstract}



\section{Introduction}\label{sec:introduction}
The latest data release, \textit{Gaia} Early Data Release 3 \citep[EDR3;][]{2016A&A...595A...1G, 2021A&A...649A...1G}, contains the astrometric data for about 1.8 billion objects. The uncertainties in parallax, sky position, and proper motion for the brightest (\textit{Gaia} $G$-band magnitude $G<15$ mag) samples are  0.02--0.03 mas, 0.01--0.02 mas, and 0.02--0.03\;$\mathrm{mas\;yr^{-1}}$, respectively. Such large amount of high quality data reveals the detailed phase-space distribution of the stars, which reflects the gravitational potential of the Milky Way (MW).

\begin{figure}
	\begin{center}
    \includegraphics[width=\columnwidth]{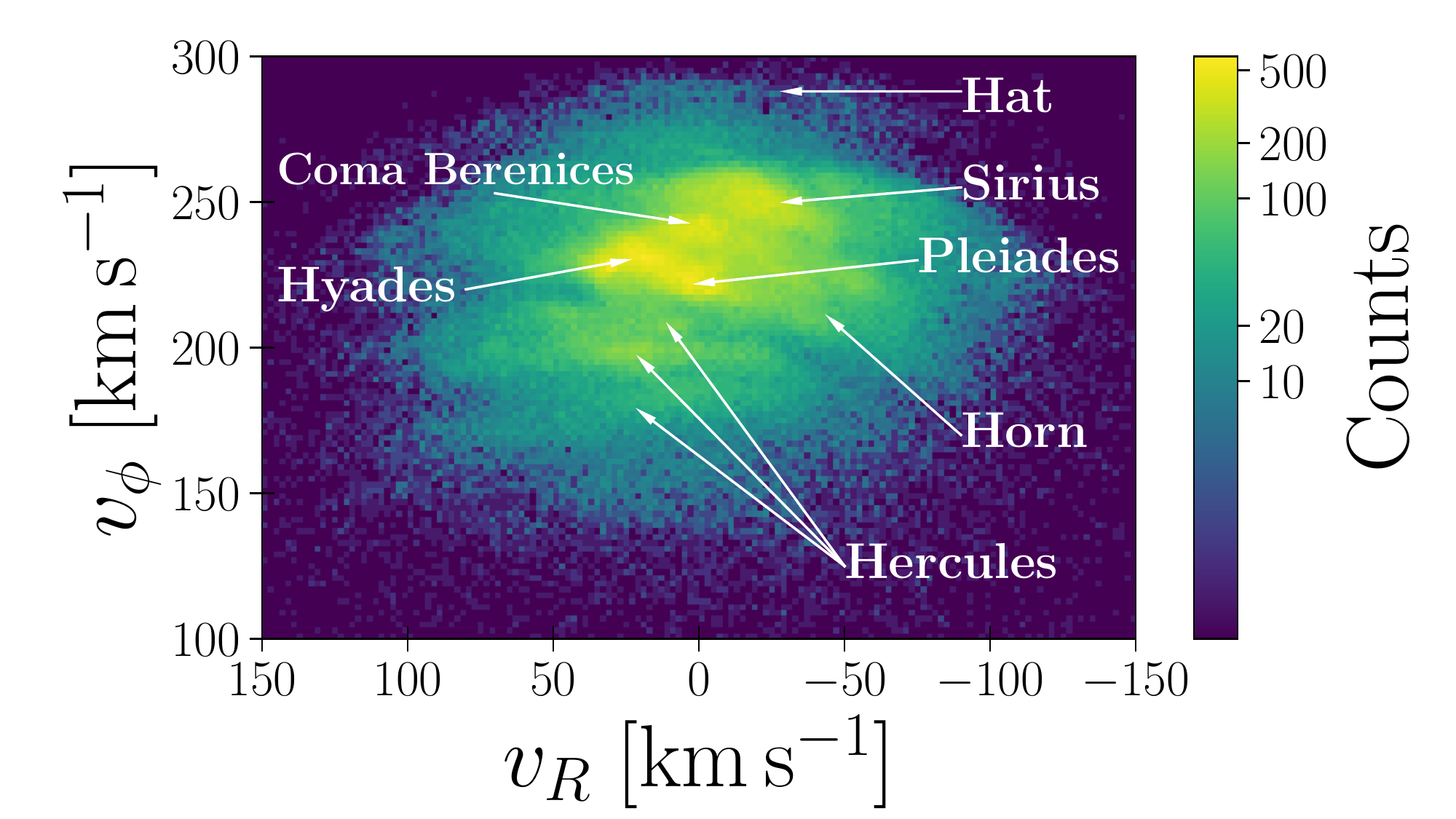}
		\caption{Solar neighbourhood star distribution in radial velocity versus azimuthal velocity ($v_R$-$v_{\phi}$) space. This plot is based on the Gaia data as described in Section~\ref{sSec:gaia_analysis}.}\label{fig:uv_gaia}
	\end{center}
\end{figure}

Fig.~\ref{fig:uv_gaia},
(created using Gaia data, see Section~\ref{sSec:gaia_analysis} for details)
shows the velocity-space distribution for the stars within 200\;pc from the Sun. We see some substructure (moving groups) in the figure \citep{2018A&A...616A..11G}. 
The names and locations of the major substructures are also shown in this figure.
The Hercules stream is one of the most prominent substructures. It was already identified from the \textit{Hipparcos} \citep{1997ESASP1200.....E, 1997A&A...323L..49P} observation \citep{1998AJ....115.2384D}.

These velocity-space substructures are not expected in axisymmetric discs, therefore their origins are often linked with Galactic non-axisymmetric structures such as the bar and the spiral arms.
\citet{1999ApJ...524L..35D, 2000AJ....119..800D} demonstrated that stars trapped in 2:1 outer Lindblad resonance (OLR) can form Hercules-like streams when using test particle simulations. In order to locate the 2:1 OLR in the solar neighbourhood, the bar's pattern speed of $\OmegaBar \gtrsim 50 \; \kmskpc$ is required \citep{1999ApJ...524L..35D, 2000AJ....119..800D, 2001A&A...373..511F, 2007ApJ...664L..31M, 2010MNRAS.407.2122M, 2014A&A...563A..60A, 2017MNRAS.465.1443M, 2017MNRAS.466L.113M, 2019MNRAS.488.3324F, 2021MNRAS.507.4409M}.
It is faster than the recently measured values of $\OmegaBar \simeq 40 \;\kmskpc$ \citep{2019MNRAS.490.4740B,2019MNRAS.488.4552S}. The measurements of the bar length \citep{2015MNRAS.450.4050W}, comparison between hydrodynamical simulations and CO and  H\,{\scriptsize I} gas observations \citep{2015MNRAS.454.1818S, 2016ApJ...824...13L, 2022ApJ...925...71L}, and dynamical models of bulge stars \citep{2017MNRAS.465.1621P, 2019MNRAS.489.3519C, 2022MNRAS.512.2171C} also support the slow bar.

Trapping in the bar's  corotation resonance (CR) is more favoured as a possible origin of the Hercules stream in the case of the slow bar.
\citet{2017ApJ...840L...2P} performed test particle simulations in a MW potential model made with made-to-measure (M2M) method \citep{2017MNRAS.465.1621P} and found that particles in CR form a Hercules-like stream.
The same scenario is proposed from $N$-body simulations \citep{2020ApJ...890..117D} and analytic models \citep{2019A&A...626A..41M, 2019A&A...632A.107M, 2020MNRAS.495..895B, 2021MNRAS.500.4710C}.
Guiding radii of higher order bar resonances like 4:1 OLR are located around the Sun's radius of $R_0 \simeq 8.2\;\kpc$ \citep{2019A&A...625L..10G} if the bar has a moderate pattern speed of $\OmegaBar \simeq40$--$45\;\kmskpc$.
They are also possible origins for the Hercules stream and the other velocity-space substructures~\citep{2017MNRAS.471.4314M, 2019MNRAS.484.4540H, 2019A&A...626A..41M, 2021MNRAS.506.4687M}.
Spiral arm resonances or their combination with bar resonances also form velocity-space substructures \citep{2018MNRAS.481.3794H, 2019MNRAS.490.1026H, 2019MNRAS.484.4540H, 2018A&A...615A..10M, 2019ApJ...876...36M, 2020ApJ...888...75B}.

Satellite galaxies such as the Sagittarius dwarf galaxy can impact the local stellar kinematics. 
\citet{2018Natur.561..360A} discovered the phase spirals (or snail shells) in $z$-$v_z$ plane. One promising scenario for this origin is that the MW disc is perturbed by the Sagittarius dwarf galaxy. In the same way, overdensities in $v_R$-$v_{\phi}$ space may also be the result of the external perturbation \citep{2019MNRAS.489.4962K, 2019MNRAS.485.3134L, 2020ApJ...890...85L, 2021MNRAS.508.1459H}.

Some scenarios for explaining the origin of the velocity-space substructures are proposed as introduced above. However, we do not have a widely accepted answer for their origin.
Some studies tackle the problem by focusing on not only the velocity space but also the other phase-space sections such as the radial action versus azimuthal action ($J_R$-$J_{\phi}$) plane \citep{2018MNRAS.477.3945H, 2019MNRAS.490.1026H, 2019MNRAS.484.3291T,2022MNRAS.509..844T, 2021MNRAS.508..728K, 2021MNRAS.500.2645T} or chemical information \citep[e.g.][]{2019MNRAS.484.4540H, 2021MNRAS.505.2412C, 2021arXiv210505263W}.

Most of the above studies use test particle simulations in static potentials and do not focus on time evolution of the potentials. 
Recently, \citet{2021MNRAS.500.4710C} and \citet{2021MNRAS.505.2412C} modelled the evolution of resonant orbits in the Galactic potential with a decelerating bar using secular perturbation theory and test particle simulations. 
The observed Hercules stream is highly asymmetric in $v_R$. Their model successfully reproduces the feature by trapping in the CR of the decelerating bar. 
Self-consistent $N$-body simulations suggest that the MW has experienced a more complex structural evolution in addition to the bar's speed slowing down \citep{1984ApJ...282...61S, 1988MNRAS.231P..25S, 2009ApJ...706..471B, 2011ApJ...730..109F, 2012MNRAS.421.1529G, 2012MNRAS.426..167G, 2013ApJ...763...46B, 2013ApJ...766...34D, 2020A&A...638A.144K, 2021arXiv211105466T}. Such a complex time evolution possibly impacts the stellar orbits and phase-space distributions.

In our previous study \citep{2020MNRAS.499.2416A}, we analysed a high-resolution $N$-body simulation of the MW and found a Hercules-like stream in the final snapshot. Orbital frequency analysis confirmed that it is made from 4:1 OLR and 5:1 OLR\@. We concluded that the trimodal structure of the Hercules stream in the MW can be explained by 4:1 OLR, 5:1 OLR, and CR in the bar's pattern speed of $\OmegaBar \simeq 40$--$45\kmskpc$. 
This study confirmed that Hercules-like streams originating from the bar resonance exist in at least one position in one snapshot.
In this paper, we analyse the same simulation data but put the focus on the time evolution and spatial variation of the local velocity-space distributions.
We use Kullback-Leibler divergence (KLD) to measure the similarity of the velocity-space distributions in the simulation and that of the solar neighbourhood stars.
In Section~\ref{sec:methods}, we briefly introduce our $N$-body model and describe the analysis. In Section~\ref{sec:results}, we show how velocity-space distributions and the KLDs vary as functions of time and spatial positions. In Section~\ref{sec:discussions}, we perform the obit analysis for the particles around the position where the velocity-space distributions with high similarities are detected. This analysis shows that velocity-space substructures such as Hercules stream are made from bar resonances. The summary of this paper is given in Section~\ref{sec:summary}.

\section{$N$-body simulations and analysis}\label{sec:methods}
\subsection{$N$-body simulations}\label{sSec:simulation}
\begin{figure*}
	\begin{center}
    \includegraphics[width=0.3\linewidth]{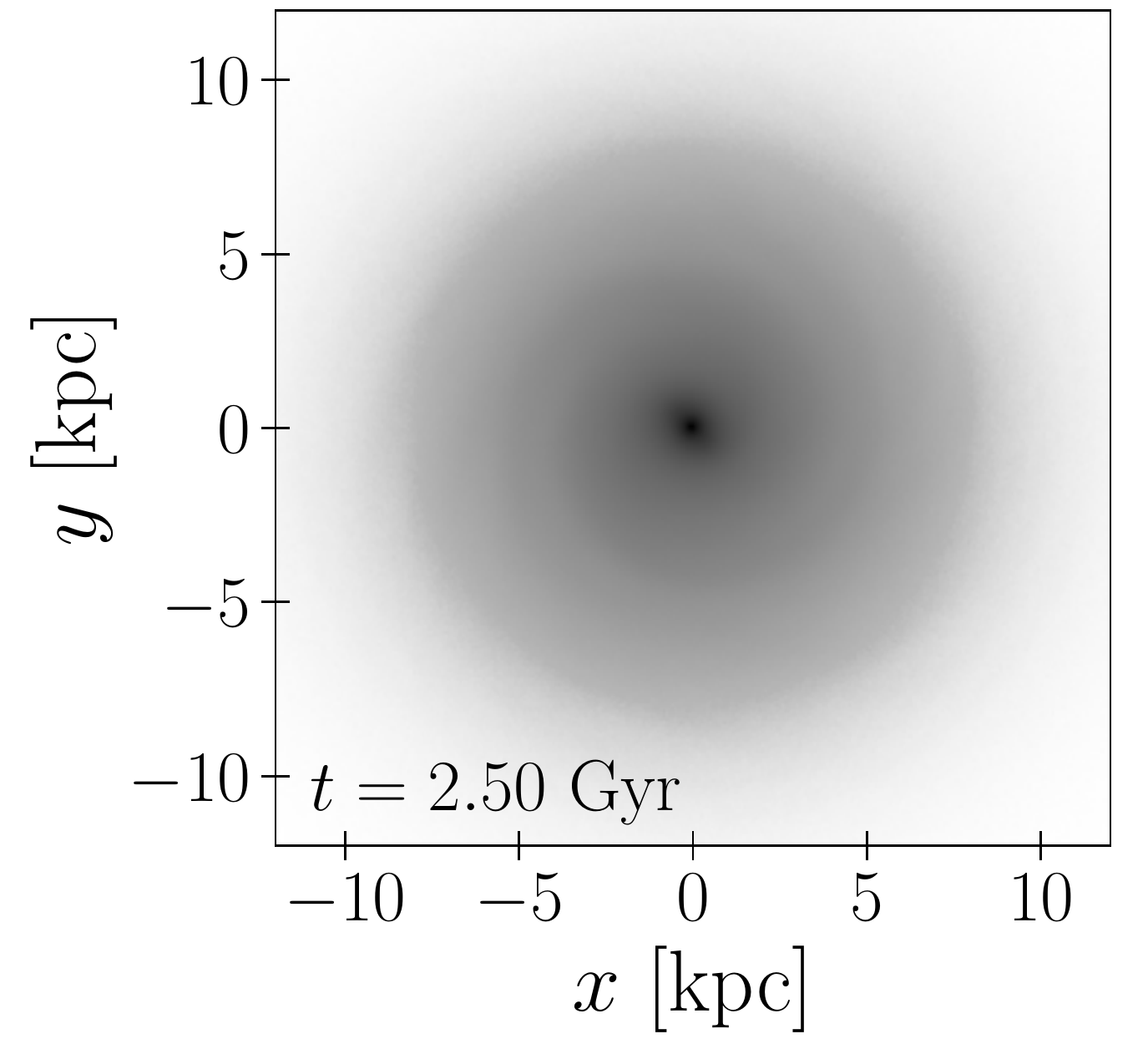}
    \includegraphics[width=0.3\linewidth]{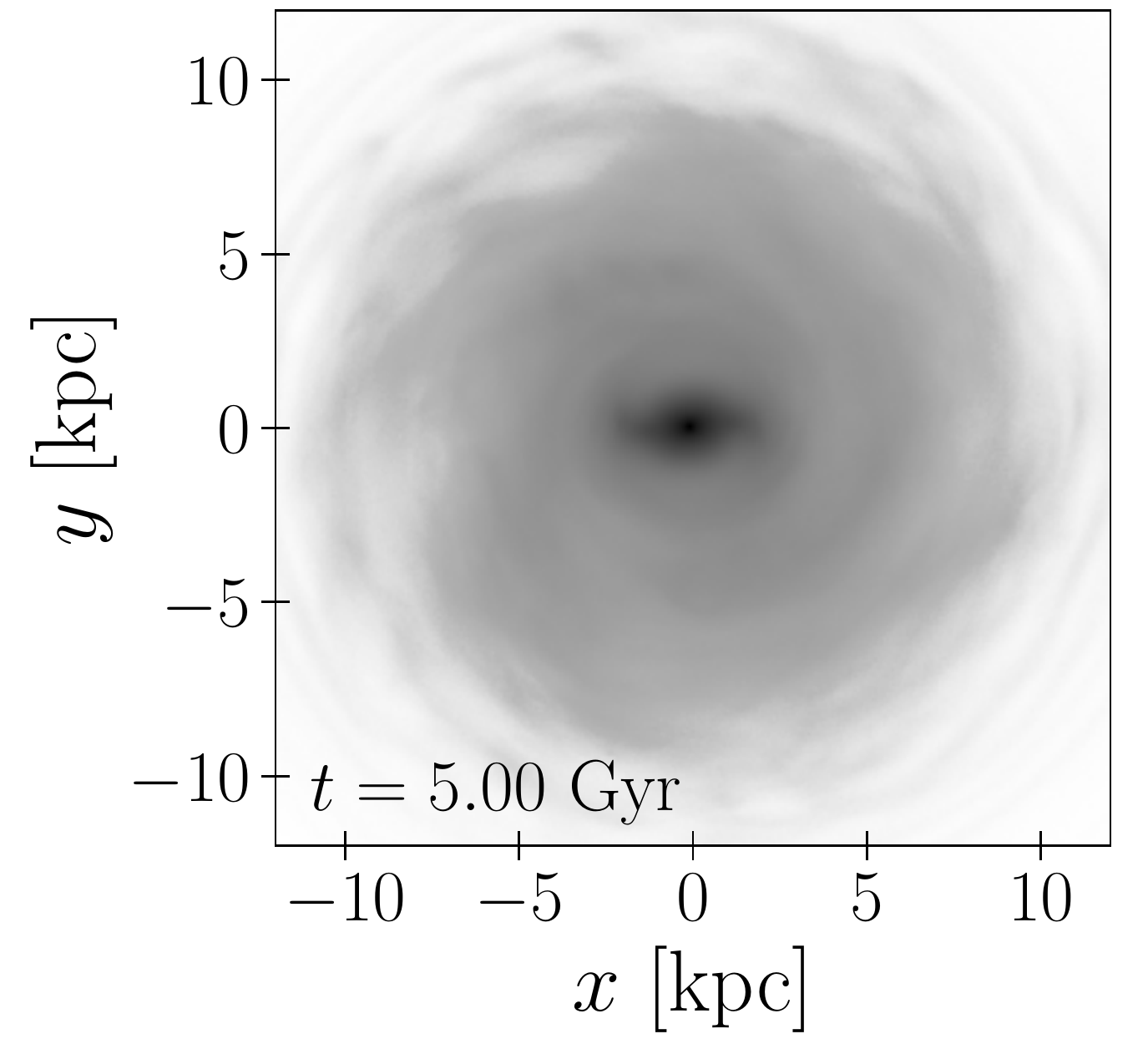}
    \includegraphics[width=0.3\linewidth]{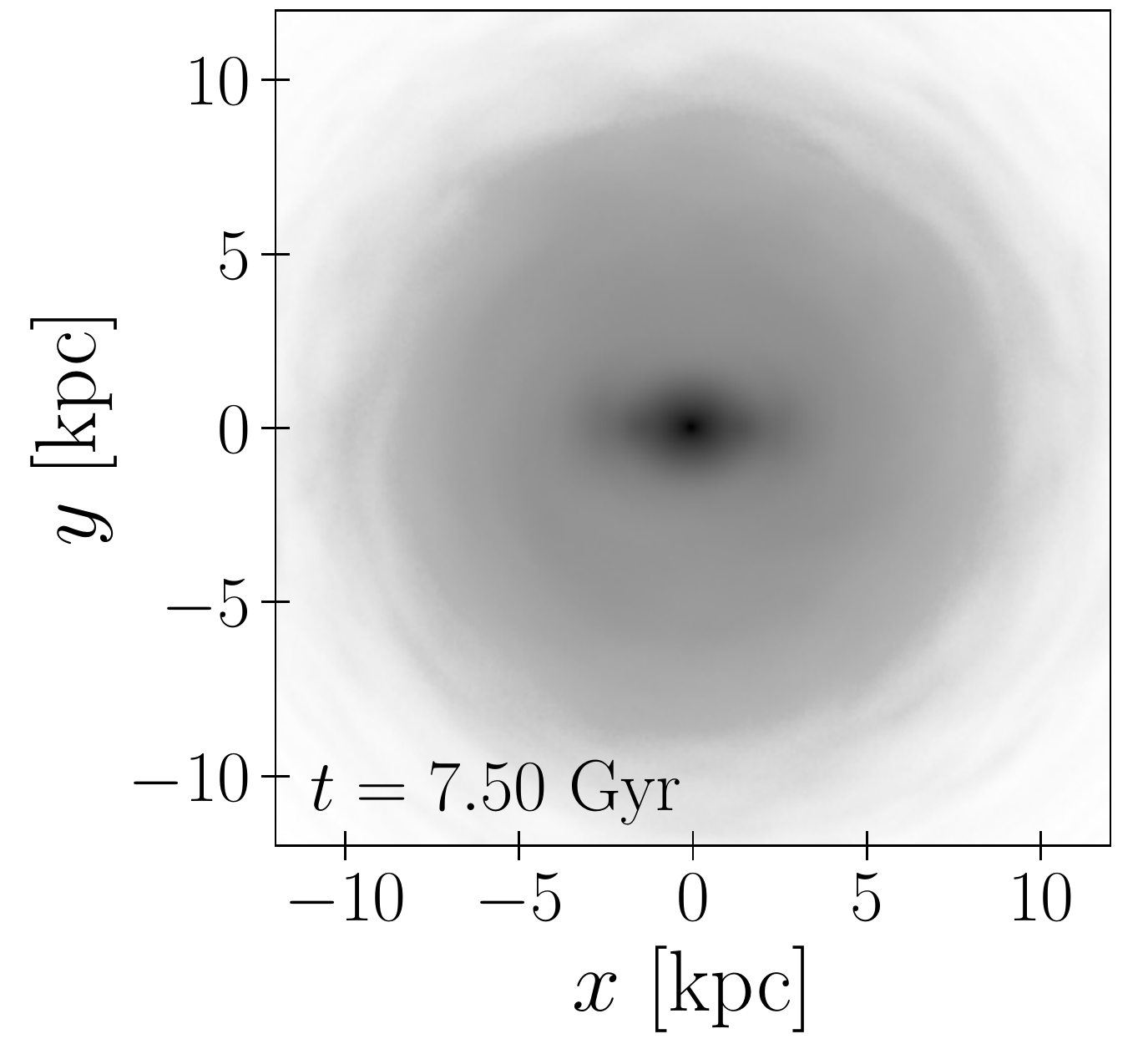}
    \includegraphics[width=0.3\linewidth]{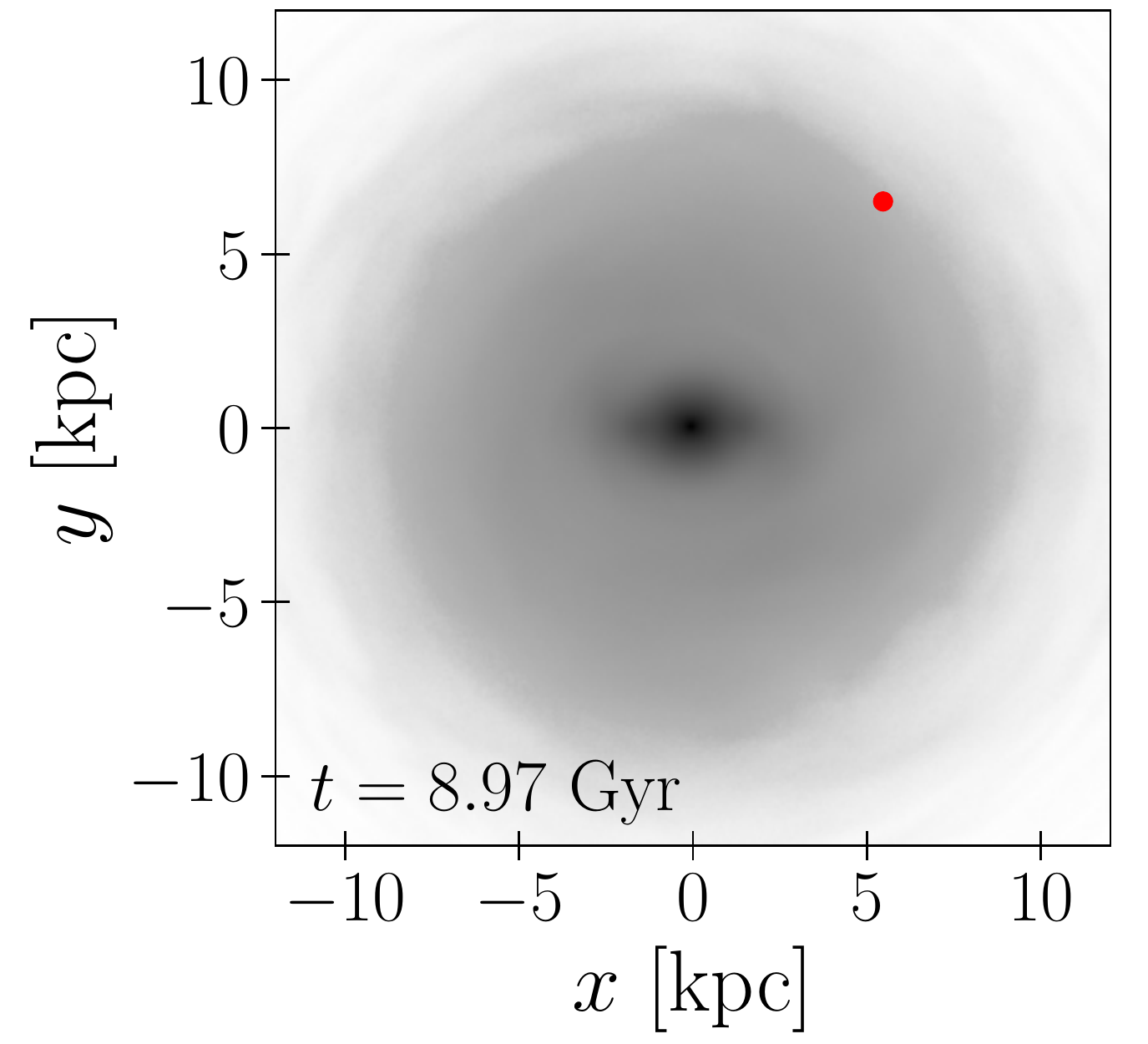}
    \includegraphics[width=0.3\linewidth]{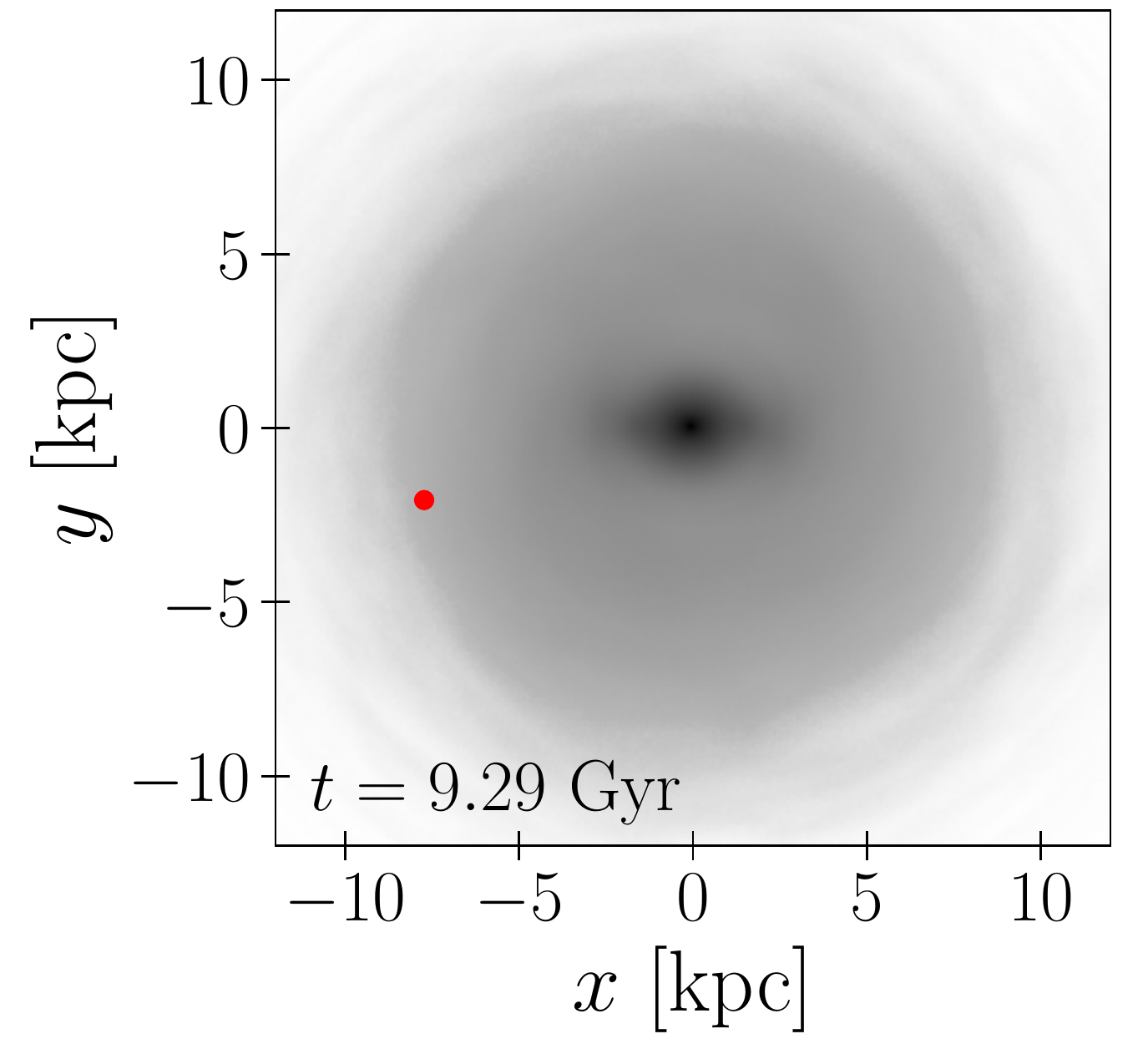}
    \includegraphics[width=0.3\linewidth]{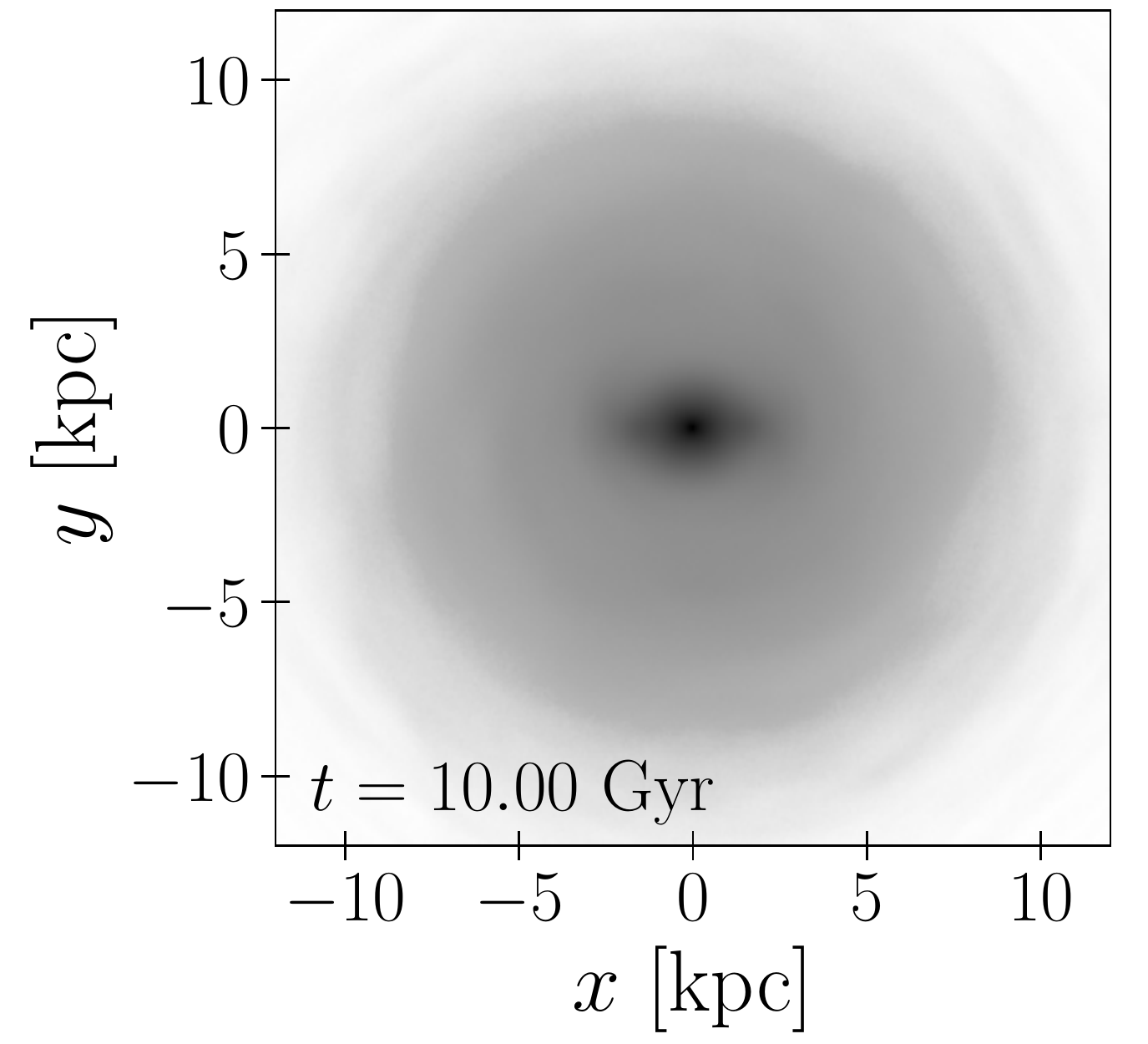}
		\caption{Face-on views of the MW model at $t=2.5$, 5, 7.5, 8.97, 9.29, and 10\;Gyr. In Section~\ref{sec:discussions}, we see the velocity-space distributions at the positions indicated by the red dots in the panels of $t=$8.97 and 9.29\;Gyr.}\label{fig:snaps_face_on}
	\end{center}
\end{figure*}

\begin{figure}
	\begin{center}
		\includegraphics[width=\columnwidth]{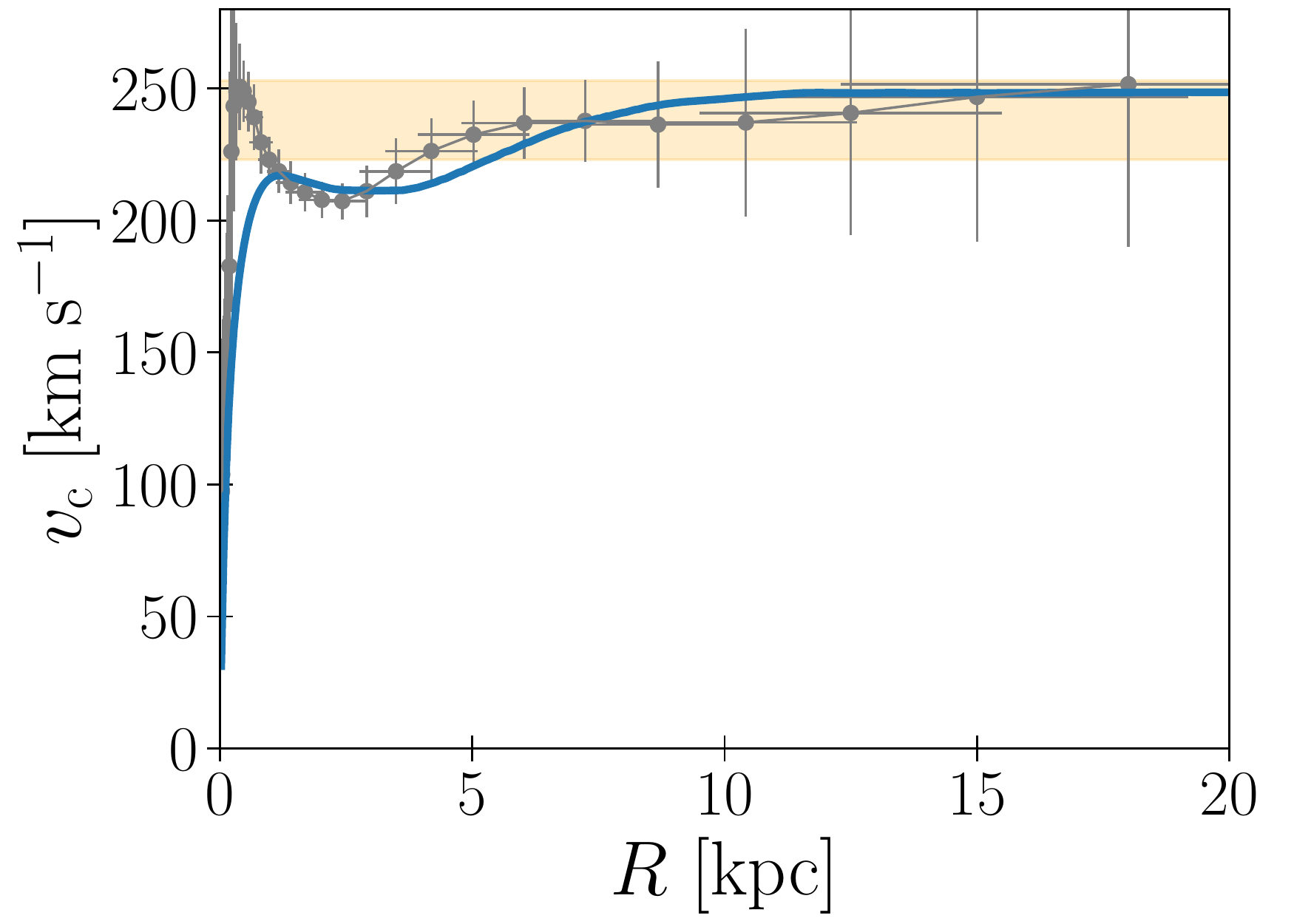}
		\caption{The rotation curves of the $N$-body model and the MW. The blue line shows the rotation curve of the $N$-body model at $t=10$.\;Gyr. The grey dots show the observed rotation curve of the MW \citep{2017PASJ...69R...1S}. The orange shaded region indicates the observationally estimated circular velocity at the Sun's radius \citep{2016ARA&A..54..529B}.}\label{fig:rot_curve}
	\end{center}
\end{figure}

\citet{2019MNRAS.482.1983F} performed $N$-body simulations of MW-like galaxies. We use one of their models, MWa.
This model is a MW-like galaxy composed of a live stellar disc, a live classical bulge, and a live dark-matter (DM) halo. The initial conditions were generated using {\tt GalactICS} \citep{1995MNRAS.277.1341K,2005ApJ...631..838W}. The stellar disc follows an exponential profile with a mass of $3.73 \times 10^{10}M_{\sun}$, an initial scale-length ($R_{\rm d}$) of $2.3$\;kpc, and an initial scale-height of $0.2$ pc. 
The classical bulge follows the Hernquist profile \citep{1990ApJ...356..359H}, whose mass and scale-length are $5.42 \times 10^9 M_{\sun}$ and $750$ pc, respectively. The DM halo follows the Navarro-Frenk-White (NFW) profile \citep{1997ApJ...490..493N}, whose mass and scale radius are $8.68 \times 10^{11}M_{\sun}$ and 10\,kpc, respectively.
The number of disc, bulge, and halo particles are 208M, 30M, and 4.9B, respectively. A more detailed model description can be found in \citet{2019MNRAS.482.1983F}.
The simulations were performed using the parallel GPU tree-code, {\tt BONSAI}\footnote{{\tt https://github.com/treecode/Bonsai}} \citep{2012JCoPh.231.2825B,2014hpcn.conf...54B} using the Piz Daint GPU supercomputer. 

The simulation was started from an axisymmetric disc without any structure at $t=0$\;Gyr and was continued up to $t=10$\;Gyr. Fig.~\ref{fig:snaps_face_on} shows the face-on views of the $N$-body model at $t=2.5$, 5, 7.5, 8.97, 9.29, and 10 Gyr. In the simulation the bar and spiral arms form spontaneously due to instabilities. The spiral structures are most prominent at $t \sim 5$\;Gyr after which they 
become fainter due to the dynamical heating of the disc.

	The simulated and observed rotation curves are plotted in Fig.~\ref{fig:rot_curve}. The grey dots with error bars are taken from \citet{2017PASJ...69R...1S}. The orange shaded region indicates the range of $v_{\mathrm{c}} = 238 \pm 14 \; \kms$, which is the circular velocity at the Sun's radius estimated by \citet{2016ARA&A..54..529B}. The $N$-body model well reproduces the other disc and bulge properties of the MW. See \citet{2019MNRAS.482.1983F} for a more detailed comparison with observations.

\begin{figure}
	\begin{center}
		\includegraphics[width=\columnwidth]{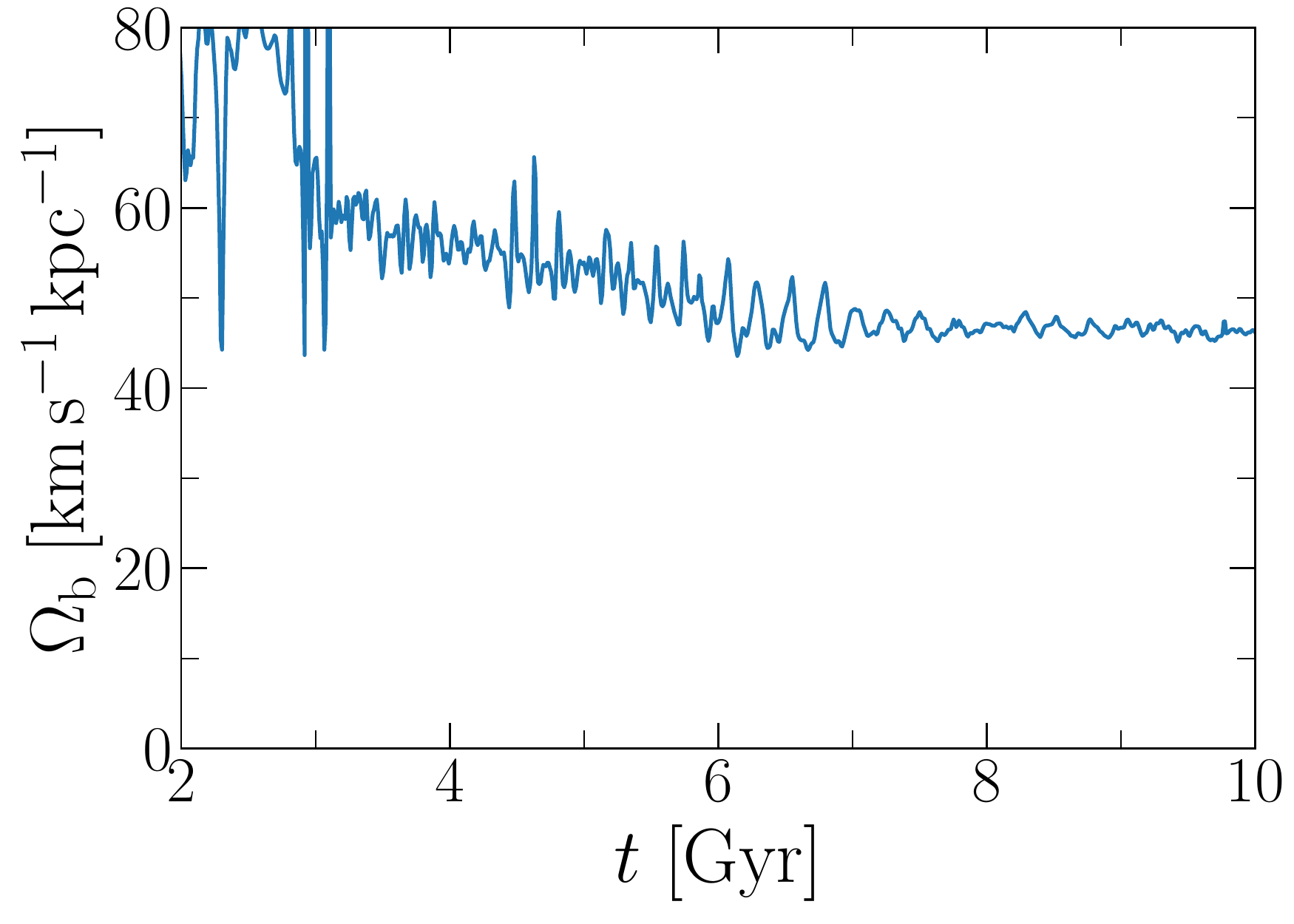}
		\caption{Time evolution of the bar's pattern speed.}\label{fig:bar_time_evolution}
	\end{center}
\end{figure}

\begin{figure}
	\begin{center}
		\includegraphics[width=\columnwidth]{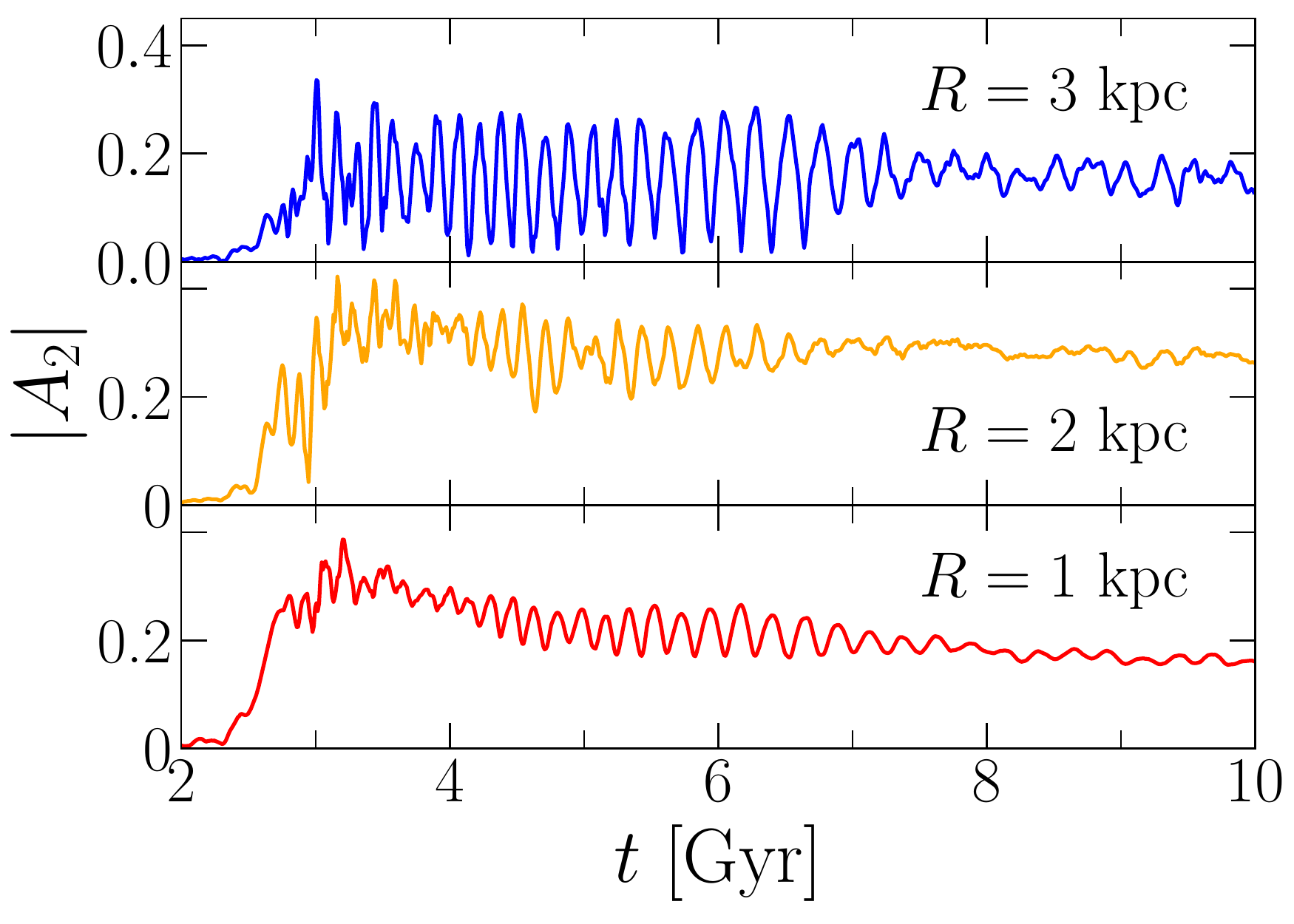}
		\caption{Time evolution of the Fourier amplitude $|A_2(R)|$ at $R=1$, 2, and 3\;kpc.}\label{fig:A2_time_evolution}
	\end{center}
\end{figure}

In this work we determine the bar's pattern speed using the Fourier decomposition as was also done in \citet{2019MNRAS.482.1983F} and \citet{2020MNRAS.499.2416A}.
We divide the galactic disc into annuli with a width of 1\;kpc, and then Fourier decompose the disc's surface density in each annulus:
\begin{equation}
	\Sigma (R, \phi) = 
	\sum_{m=0}^{\infty} A_m(R) \exp\{im[\phi - \phi_m(R)]\},
	\label{eq:Fourier_Decomp}
\end{equation}
where $A_m(R)$ and $\phi_m(R)$ are the $m$-th mode's amplitude and phase angle, respectively.  
We define $\phi_2(R)$ averaged in $R<3$\;kpc as the angle of the bar in the snapshot \citep[see also][]{2019MNRAS.482.1983F}. 
The bar's pattern speed, $\OmegaBar$, is determined using the least squares fitting to $\phi_2 (t) = \OmegaBar t + \phi_{2, 0}$.
Fig.~\ref{fig:bar_time_evolution} and Fig.~\ref{fig:A2_time_evolution} show the time evolution of the pattern speed and the Fourier amplitude $|A_2|$, respectively.
One can see that after $\sim 2$\,Gyr, a bar started to form, and continued to grow until $\sim 5$\,Gyr (Fig.~\ref{fig:A2_time_evolution}). During the evolution, the bar slowed down with oscillations up to $\sim 7$\,Gyr (Fig.~\ref{fig:bar_time_evolution}). 

\subsection{Orbit analysis}\label{sSec:orbit_analysis}
Here, we summarise the orbit analysis method used to select particles trapped in resonances.
A more detailed descriptions can be found in \citet{2020MNRAS.499.2416A}.

Resonantly trapped particles are selected via their orbital frequency ratios. We follow \citet{2007MNRAS.379.1155C}'s method to compute orbital frequencies.
We determine the radial frequency, $\Omega_R$, using the Discrete Fourier Transformation (DFT) for $R(i)$ where $R(i) \; (i=1,\ldots,64)$ is a radial coordinate in the $i$-th snapshot.
We employ a zero-padding technique of Fourier transforming:\  960 zero points are added at the end of the data series.
We then sample the frequency space with 512 points between $0\; \kmskpc$ and
$315\;\kmskpc$, whereby the upper bound is given by the Nyquist frequency.
We identify $\Omega_R$ as a frequency that causes a local maximum in the Fourier spectrum.
In contrast, the associated angular frequency $\Omega_{\phi}$ is determined via least-squares fitting instead of DFT.\@
From the snapshots, we collect, per particle, the series of measured angles $\phi(i)$ as a function of time $t(i)$, where $i$ iterates over the 64 snapshots available. For each particle, this results in pairs $[t(i), \phi(i)]\; (i=1,\ldots,64)$ to which we fit the function $\phi = \Omega_{\phi} t +\phi_0$ using a least squares method. 

The resonance condition for $m$:$l$ OLR are represented as $m(\Omega_{\phi}-\OmegaBar) + l\Omega_R = 0$. Orbits trapped in the resonance distribute around $-l/m$ in the frequency ration ($(\Omega_{\phi}-\OmegaBar)/\Omega_R$) space. 
We select particles whose frequency ratios are within a range of $\pm0.01$ from the exact resonance frequency ratio as particles trapped in the resonance. As demonstrated in \citet{2020MNRAS.499.2416A}, this procedure selects resonantly trapped particles without contamination.

\subsection{Analysis of the \textit{Gaia} data}\label{sSec:gaia_analysis}
From the \textit{Gaia} EDR3 catalogue, we select samples that satisfy (1) a relative parallax error of less than 20\% ($\varpi/\sigma_{\varpi} > 5$), (2) the distance from the Sun is less than 200\;pc ($1/\varpi [\mathrm{mas}] < 0.2$), and (3) the radial velocity error is less than $5\;\kms$.
We use \texttt{astropy.coordinates} from the \texttt{astropy} \citep{2013A&A...558A..33A} Python package to convert from heliocentric to Galactocentric coordinates.
We assume that the distance of the Sun from the Galactic centre is $R_0 = 8.178$\;kpc \citep{2019A&A...625L..10G}, and that the distance of the Sun from the Galactic mid-plane is $z_0 = 20.8$\;pc \citep{2019MNRAS.482.1417B}, whereby the velocity of the Sun with respect to the local standard of rest (LSR) is $(U_{\sun}, V_{\sun}, W_{\sun}) = (11.1, 12.24, 7.25)\;\kms$ \citep{2010MNRAS.403.1829S}, and the azimuthal velocity of $\Theta_{\sun} + V_{\sun} = 247.4\;\kms$ \citep{2004ApJ...616..872R, 2019A&A...625L..10G}.

\subsection{Kullback-Leibler divergence}\label{sSec:KLD_analysis}
We use Kullback-Leibler divergence \citep[KLD;][]{kullback1951information} to quantitatively evaluate the similarity of the velocity-space distributions in the simulation and that in the observation.
KLD is defined between two probability distributions. When $p(x)$ and $q(x)$ are discrete distributions in a probability space $\mathcal{X}$, the KLD between them is defined as
\begin{equation}
	D(p||q) = \sum_{x\in \mathcal{X}} p(x)\log \frac{p(x)}{q(x)}.
	\label{eq:KLD}
\end{equation}
KLD satisfies the following properties like a `distance' between $p$ and $q$.
\begin{enumerate}
	\item $D(p||q) \ge 0$.
	\item $D(p||q) = 0$ if and only if $p = q$.
\end{enumerate}
We note that $D(p||q)$ is not equal to $D(q||p)$, hence it is not a distance in a mathematical sense. To be more precise, $D(p||q)$ represents how a distribution $q$ differs from a reference distribution $p$.
In this study we would like to evaluate how well the velocity-space distributions in the simulation reproduce that of the MW, hence $p$ and $q$ are determined from the observation and the simulation, respectively. 
The distribution $p$ is determined from the \textit{Gaia} EDR3 data \citep{2021A&A...649A...1G}.
We first divide the velocity space in a grid shape to convert the observation and simulation data to probability distributions. 
There are some uncertainties in the rotation curves of the MW and the Sun's velocity with respect to the LSR \citep{2016ARA&A..54..529B}. 
To take the uncertainties into account, we use relative velocities, $\hat{v}_R = (v_R - \overline{v}_R)/\overline{v}_{\phi}$ and $\hat{v}_{\phi} = (v_{\phi} - \overline{v}_{\phi})/\overline{v}_{\phi}$, instead of absolute values. Mean velocities $(\overline{v}_R, \overline{v}_{\phi})$ are determined for the stars within 200\;pc from the Sun.
We divide the $\hat{v}_R$ versus $\hat{v}_{\phi}$ space in a range of $(-0.5,0.5) \times (-0.333,0.333)$ into 48 $\times$ 32 cells.
We determine the probability that we find a star in a cell, dividing the star count in the cell by the total number of stars in all the cells.
Eq.~(\ref{eq:KLD}) indicates that KLD can diverge if a probability is zero in a cell (i.e.\ it does not contain stars).
In order to avoid that, we alternatively use the kernel density estimation (via \texttt{scipy.stats.gaussian\_kde} \citep{2020NatMe..17..261V}) as value for the  empty cells. 
The value in a cell at $(\hat{v}_R,\hat{v}_{\phi})$ is represented as
\begin{align}
	f(\hat{v}) = \sum_{i=1}^N \prod_{j=\{R,\phi\}}\frac{w_j}{h_j}K\left( \frac{\hat{v}_j-\hat{v}_{ij}}{h_j}\right); 
	\quad K(x) = \frac{1}{\sqrt{2\pi}}\exp\left( - \frac{x^2}{2}\right).
\end{align}
$N$ is the total number of the stars in the sampling volume. $\hat{\bm{v}}_i$ is the relative velocity of the $i$-th star. $w_R$ and $w_{\phi}$ are the cell widths with $w_R = w_{\phi} = 0.0208$.
Kernel widths $h_R$ and $h_{\phi}$ are determined with the method described in \citet{scott2015multivariate}.
The typical values of $h_R$ and $h_{\phi}$ are 0.03 and 0.02, respectively.

\section{Results}\label{sec:results}
\subsection{Time evolution of the KLDs}\label{sSec:time_evolution}
\begin{figure*}
	\begin{center}
		\includegraphics[width=\columnwidth]{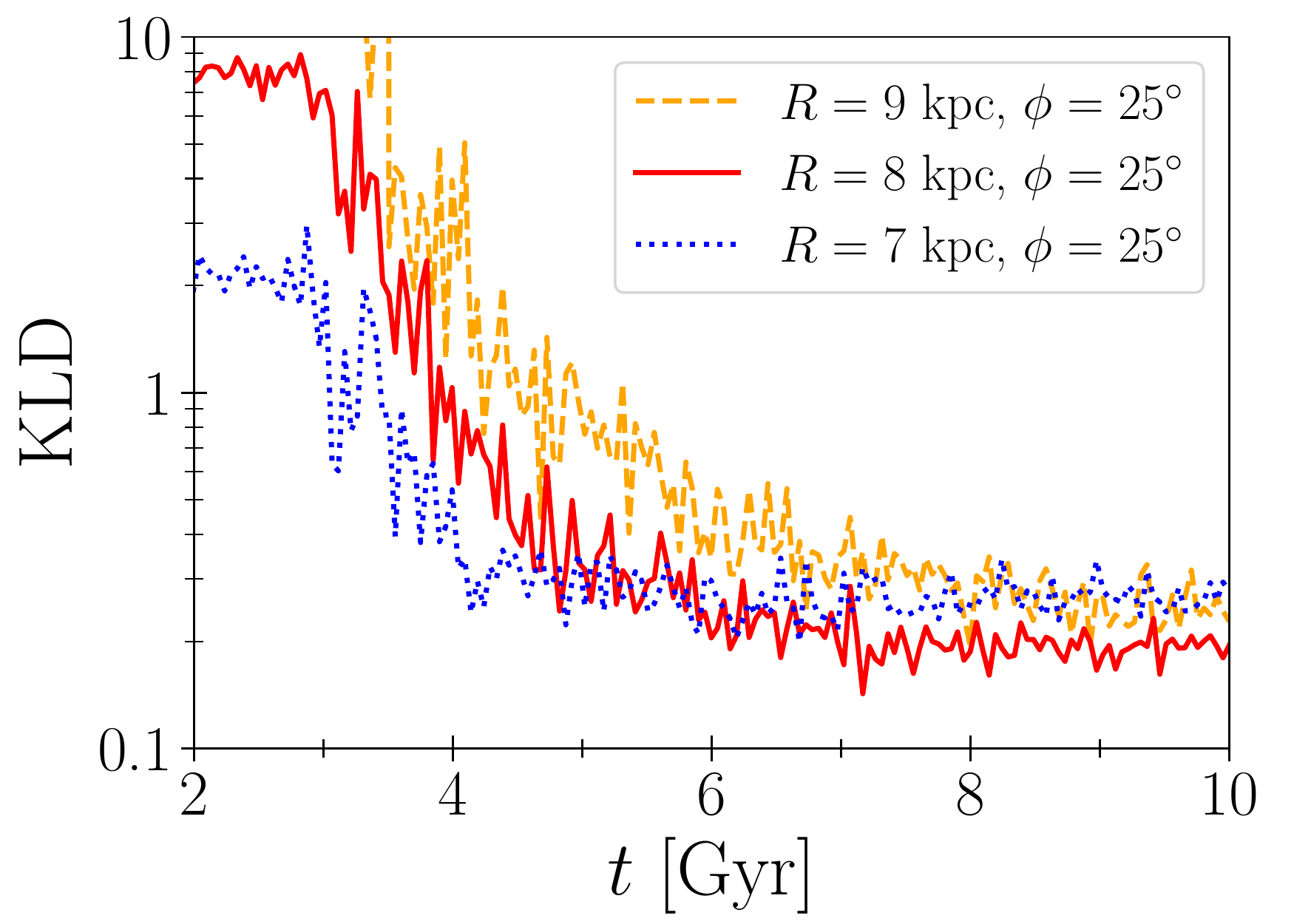}
		\includegraphics[width=\columnwidth]{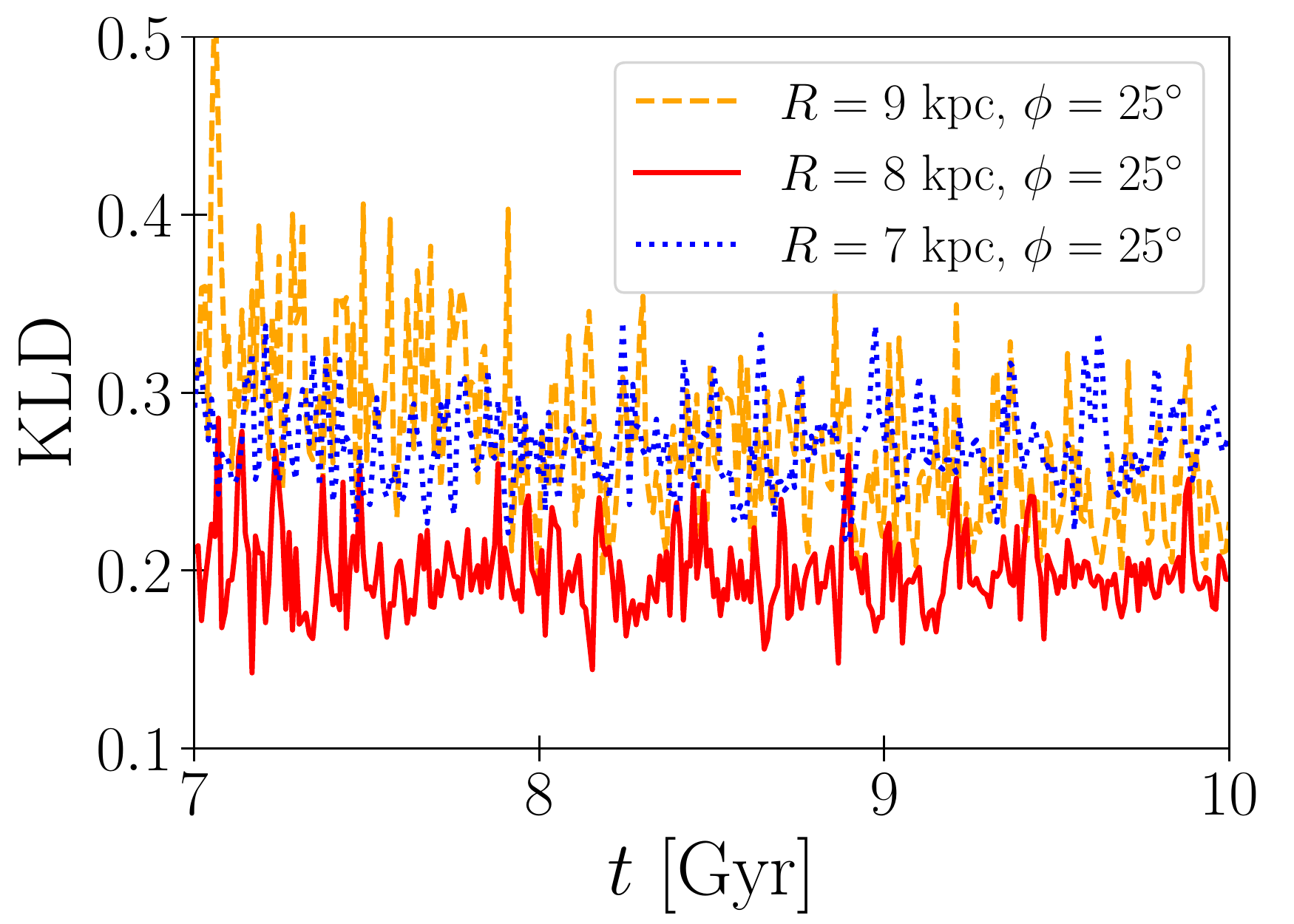}
		\caption{\textit{Left:} Time evolution of the KLDs at fixed positions in the simulation. Blue, red, and yellow lines show the KLDs at $(R,\phi) = (7\;\kpc, 25\degr)$, $(8\;\kpc, 25\degr)$, and $(9\;\kpc, 25\degr)$, respectively. \textit{Right:} Same as the left panel but for the later  time period}\label{fig:KLD_time_evolution}
	\end{center}
\end{figure*}

In this section, we investigate the time evolution of the KLDs and velocity-space distributions in the simulation.
We evaluate KLDs between the velocity-space distribution for the stars within 200~pc from the Sun and those for particles within 200~pc from $(R,\phi) = (7\;\kpc, 25\degr)$, $(8\;\kpc, 25\degr)$, and $(9\;\kpc, 25\degr)$ in the simulation, where $R$ and $\phi$ are the distance from the galactic centre and the angle with respect to the bar's major axis, respectively. 
We assume that the `Sun' in the simulation locates on the galactic mid-plane ($z=0$).
Fig.~\ref{fig:KLD_time_evolution} shows the KLDs at the three points as a function of time for both the long-term and short-term evolution. On a long time scale they decrease with time, and on a short time scale they oscillate.
The long-term evolution indicates that the velocity-space distributions are more similar to that in the observed solar neighbourhood during the later epochs than in the early epochs of the simulation.
This is a natural consequence of the setup of the simulation.
\citet{2019MNRAS.482.1983F} adjusted the initial conditions of the simulation so that the final snapshot fits the observations.
The short-term evolution indicates that the similarity of the velocity-space distribution fluctuates rapidly. Even if we find a velocity-space distribution similar to the observed one at a certain position at a certain time, a velocity-space distribution at the same position at another time is not necessarily similar to the observed one.

We can identify the correlation between the time evolutions of the KLDs and the time evolution of the bar by comparing Fig.~\ref{fig:bar_time_evolution} 
and Fig.~\ref{fig:KLD_time_evolution}. A clear bar structure appears at $t\simeq3$\;Gyr from the beginning of the simulation. The KLDs start decreasing at this time. 
From $t\simeq3$\;Gyr to $t\simeq7$\;Gyr, the bar's pattern speed slows down. During this phase, the KLDs also decrease with time.
The bar is more stable after $t \simeq 7$\;Gyr than before that although the pattern speed shows small fluctuations. In this epoch, the KLDs at $(R,\phi)=(7\;\kpc, 25\degr)$ and $(8\;\kpc, 25\degr)$ do not evolve monotonously but fluctuate around 0.3 and 0.2, respectively. The KLD at $(9\;\kpc, 25\degr)$ decreases slowly.
The bar's pattern speed is a key parameter in discussions on bar resonances. It determines the resonance radii when azimuthal and radial frequencies are given as functions of radial coordinate $R$ \citep{2008gady.book.....B}. This correlation between the KLDs and the bar's pattern speed implies that the bar resonances play an important role in regulating the local velocity-space distributions.
In Section~\ref{sec:discussions}, we discuss the relation between the velocity-space substructures and bar resonances.

The fluctuations of the KLDs are also important. Although the  KLD at $(R,\phi) = (8\;\kpc, 25\degr)$ is smaller than those at the other two positions at $t \gtrsim 7$\;Gyr, it fluctuates with time. We do not always observe the velocity-space distributions similar to that in the solar neighbourhood at this position. 
This emphasizes the non-static nature of the galaxy in the simulation.

\subsection{Angle with respect to the bar and spirals}\label{sSec:angles}
\subsubsection{Angle with respect to the bar}

\begin{figure}
	\begin{center}
		\includegraphics[width=\columnwidth]{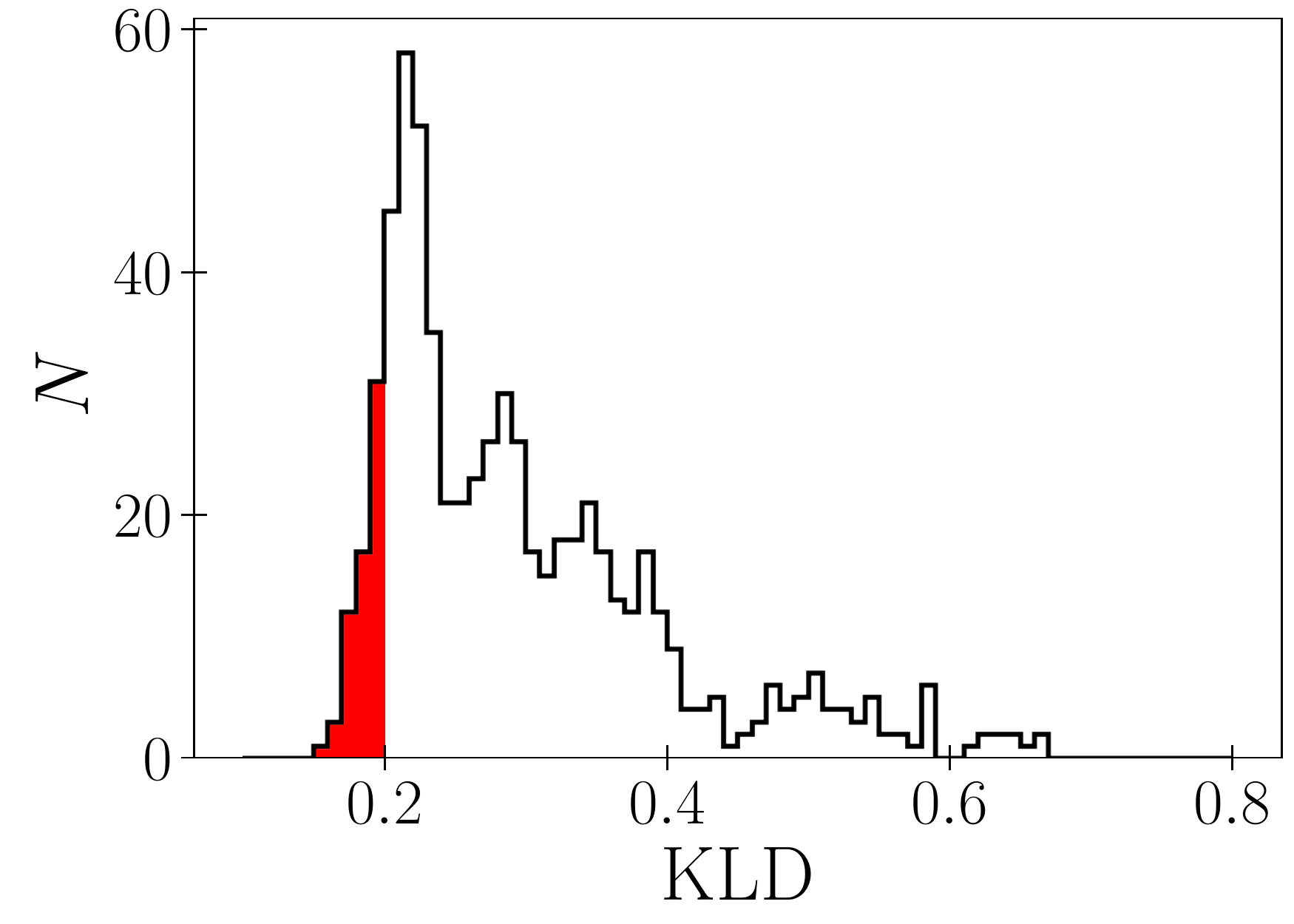}
		\caption{Distribution of the KLDs at $t=10$\;Gyr. The black line shows the histogram of the KLDs at all of the positions where the KLDs are computed. The region of $\mathrm{KLD}<0.2$ is filled with red.}\label{fig:KLD_hist_1024}
	\end{center}
\end{figure}

\begin{figure}
	\begin{center}
		\includegraphics[width=\columnwidth]{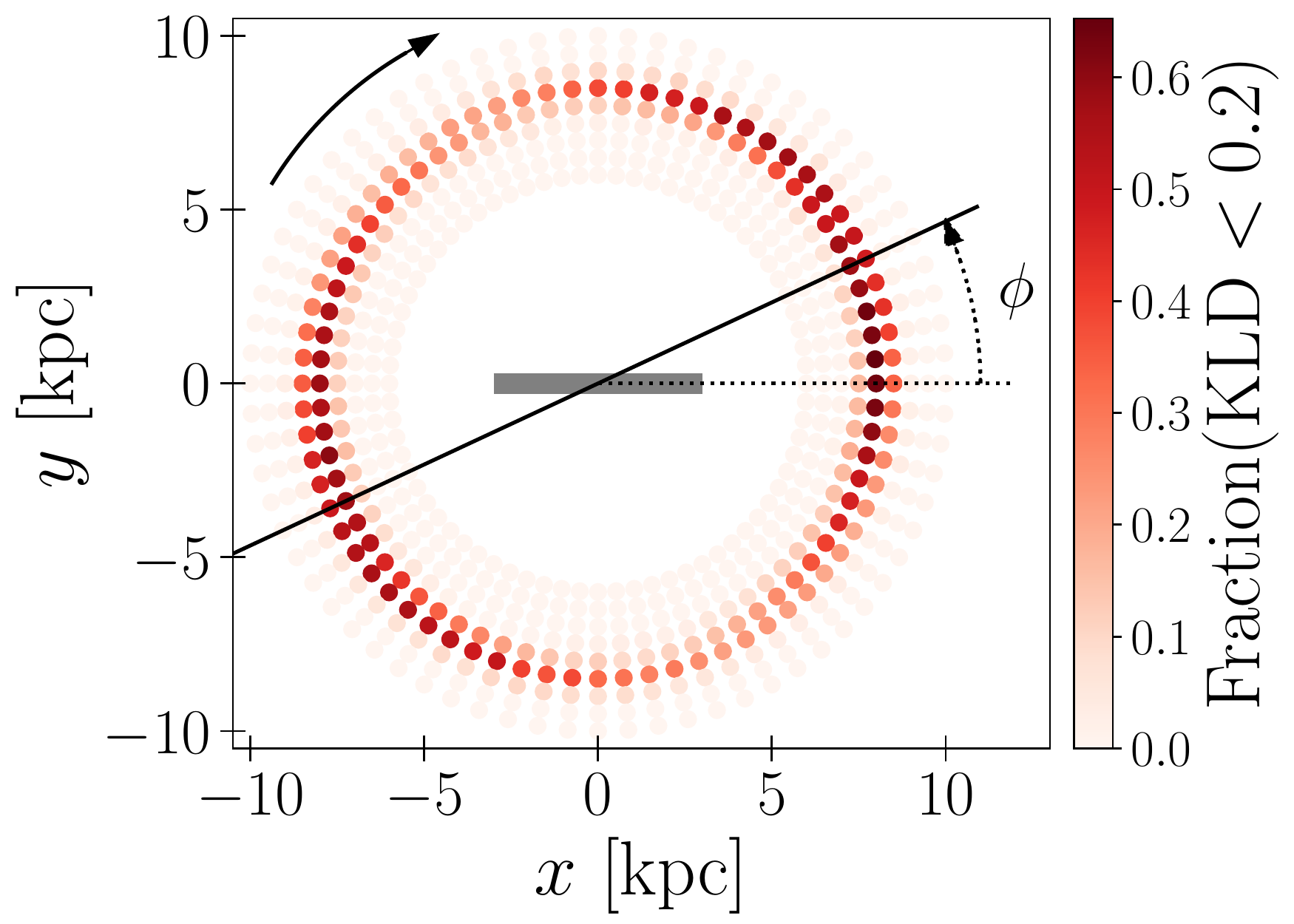}
		\caption{Positions where the velocity maps match that of the solar neighbourhood. The colours of the points represent the fractions of the number of times where KLDs less than 0.2 are detected at each position. The horizontal line and the solid arrow represent the bar orientation and the direction of the galaxy rotation, respectively. The black line represents the angle with respect to the bar of $\phi = 25^{\circ}$}\label{fig:face_on}
	\end{center}
\end{figure}

In the previous section we saw that  after $t\simeq7$\;Gyr the KLD at $(R,\phi) = (8\;\kpc, 25^{\circ})$ tends to be smaller than at $(R,\phi) = (7\;\kpc, 25^{\circ})$ and $(R,\phi) = (9\;\kpc, 25^{\circ})$.
Here, we investigate where in the disc we often detect velocity-space distributions similar to that in the solar neighbourhood.
We evaluate the KLDs of the velocity-space distributions at 648 points in the disc in the snapshots after $t=7$\;Gyr. Their positions in the galactocentric cylindrical coordinate are
\begin{align}
	\left\{
		\begin{aligned}
			R &= 6\;\kpc +\Delta R \times i \quad (i = 0,\ldots,8),  \\
			\phi &= -180^{\circ} + \Delta \phi \times j \quad (j = 0,\ldots,71),\\
			z &= 0\;\kpc,
		\end{aligned}
	\right.
\end{align}
where $\Delta R = 0.5\;\mathrm{kpc}$ and $\Delta \phi = 5^{\circ}$.
We determine the velocity-space distribution for particles within 200~pc from each of the points and compute the KLD.\@
We define that the velocity-space distribution in the simulation is similar to that in the solar neighbourhood if its KLD is less than 0.2.
We select the threshold of 0.2 because the KLD of the velocity-space distribution for the particles within 200~pc from $(R,\phi)=(8\;\kpc,20^{\circ})$ at $t=10$\;Gyr is $\sim0.2$. The velocity map is the one which we judged, by eye, to be similar to the map in the observation in our previous study \citep{2020MNRAS.499.2416A}. Fig.~\ref{fig:KLD_hist_1024} shows the distribution of the KLDs at $t=10$ Gyr. The velocity-space distributions whose KLDs lie in the red filled region ($\mathrm{KLD}<0.2$) have sufficiently high similarities.

Fig.~\ref{fig:face_on} shows the number of the times that the KLDs less than 0.2 are detected at each position. Velocity-space distributions with these small KLDs are detected more frequently at $R=8$\;kpc and $8.5$\;kpc than the other radii. Especially around $(R,\phi) = (8\;\kpc,20^{\circ})$ and $(R,\phi) = (8.5\;\kpc,50^{\circ})$, the KLDs are smaller than 0.2 for more than 50\% of the analysed snapshots.
On the other hand, at $R  \lesssim 7\;\kpc$ or $R \gtrsim 9\;\kpc$ the KLDs are larger than 0.2 for almost every snapshot.

\citet{2019MNRAS.482.1983F} obtained similar results using a simpler analysis method. They fitted the sum of two Gaussian functions with the particle distributions in $v_R$ and detected two-peak (i.e.\ Hercules-like) features.
Hercules-like features do not always appear at a fixed position.
The detection frequency is at most 50\% around $R \simeq 9$\;kpc, which is slightly outside the 2:1 OLR radius.
We only seldom detect velocity-space distributions similar to that in the solar neighbourhood around $R \simeq 9$\;kpc. The difference may be due to that \citet{2019MNRAS.482.1983F} focused only on one dimensional velocity ($v_R$) distributions.

\begin{figure}
	\begin{center}
		\includegraphics[width=\columnwidth]{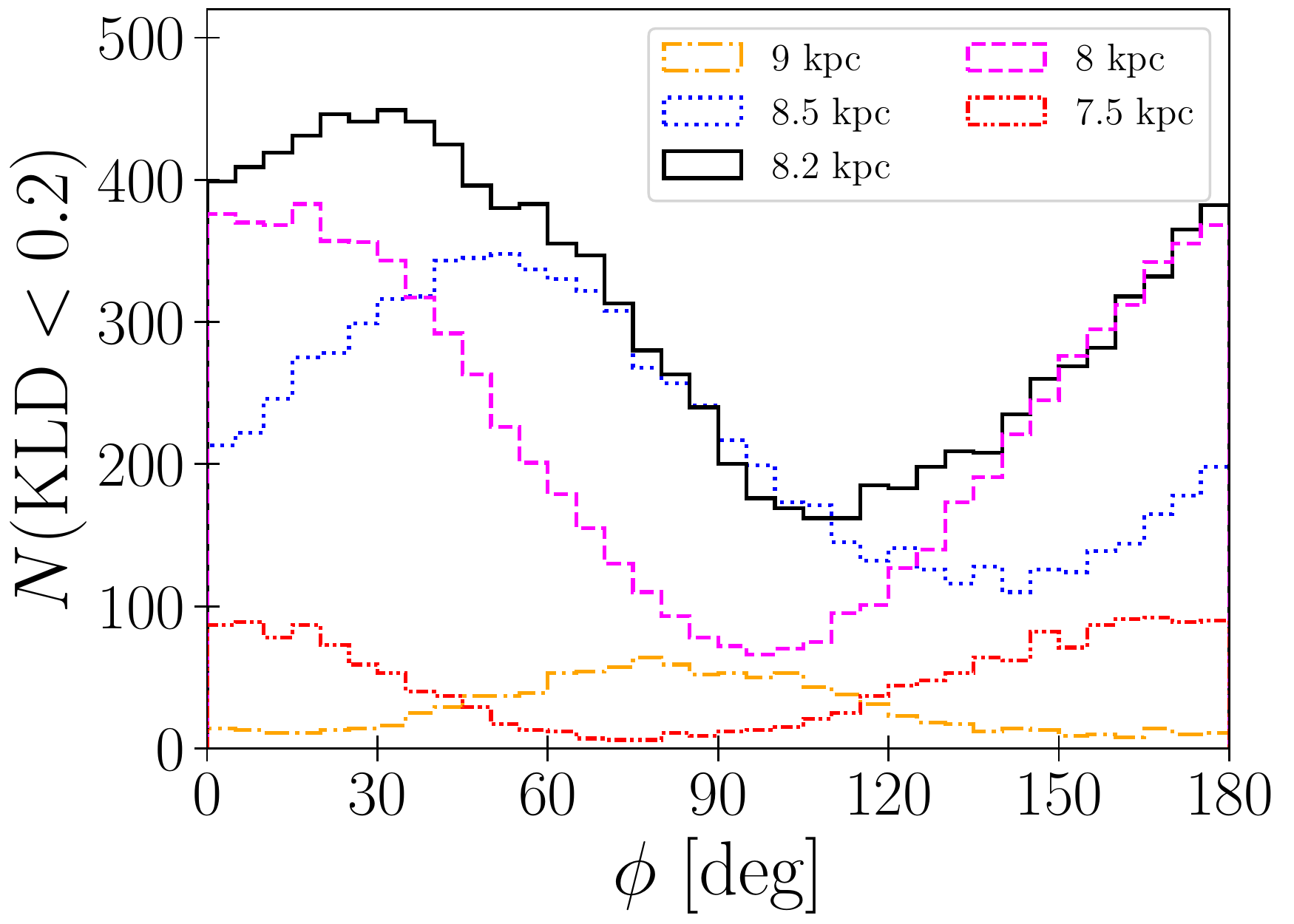}
		\caption{Histograms of the angles where the KLD of velocity distribution is less than 0.2. The angle $\phi$ is with respect to the bar. The red, magenta, black, blue, and yellow lines show the histograms at $R=7.5$\;kpc, 8\;kpc, 8.2\;kpc, 8.5\;kpc, and 9\;kpc, respectively.}\label{fig:hist_bar_angle}
	\end{center}
\end{figure}

Fig~\ref{fig:hist_bar_angle} shows the $\phi$ dependence of the KLD more clearly. In the figure, we plot the histograms for the angles of positions where the KLDs are less than 0.2 at $R=$7.5\;kpc, 8\;kpc, 8.2\;kpc, 8.5\;kpc, and 9\;kpc. 
As already seen in Fig.~\ref{fig:face_on}, these small KLDs are more frequently detected at $R=$8--8.5\;kpc. These values are close to distance between the Sun and the Galactic centre \citep{2016ARA&A..54..529B}. 
The peaks of the histograms differ by $R$.
The peak moves in the positive direction of $\phi$ as $R$ increases.
The peak of the histogram at $R=8.2\;\kpc$ locates at $\phi\simeq30\degr$, which is consistent with observationally suggested bar angle \citep{2002MNRAS.330..591B, 2007MNRAS.378.1064R, 2013MNRAS.434..595C, 2013MNRAS.435.1874W}.
The $R$ and $\phi$ dependence of the KLD also implies that the velocity-space distributions are related to bar resonances.
Particles trapped in bar resonances do not distribute uniformly in the disc, instead their distributions are dependent on $R$ and $\phi$. \citep{2007MNRAS.379.1155C, 2020A&A...634L...8K}.

\begin{figure}
	\begin{center}
		\includegraphics[width=\columnwidth]{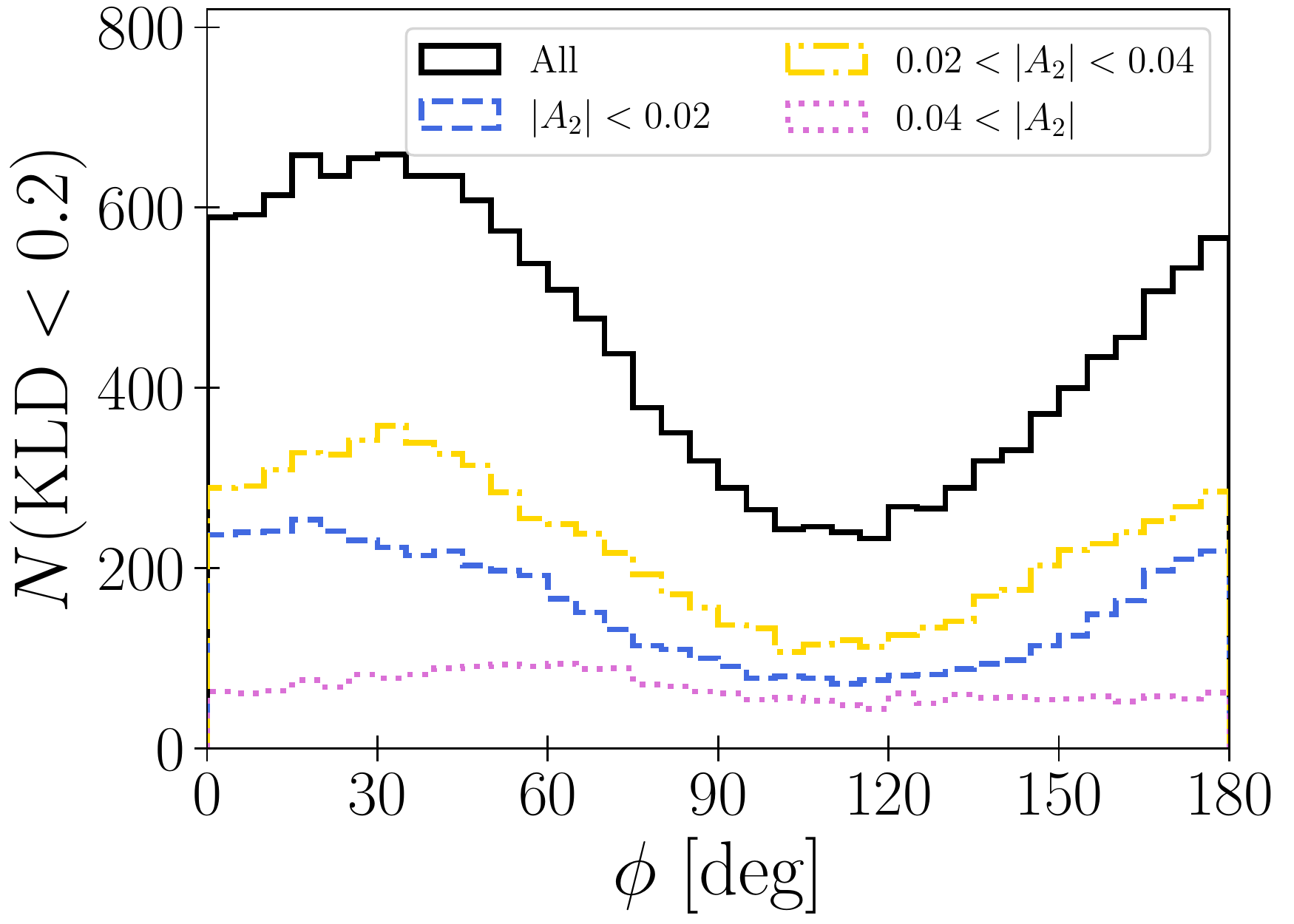}
		\caption{Histograms of the angles where the KLD of velocity distribution is less than 0.2. Blue, yellow, and magenta lines show the histograms in the cases of weak ($|A_2|<0.02$), intermediate ($0.02<|A_2|<0.04$), and strong ($0.04 <|A_2|$) spirals, respectively. The black line shows the histogram for all the spiral strength.}\label{fig:hist_bar_angle_A2}
	\end{center}
\end{figure}

\subsubsection{Angle with respect to the spirals}
Not only the bar but also the spiral arms have impact on the stellar distribution and KLD.\@  We define the spiral position as a phase angle of Fourier $m = 2$ mode $\phi_2(R )$ and define the spiral strength as a Fourier amplitude $|A_2(R )|$.
In Appendix~\ref{sec:spiral_positions}, the phase angles of $m=$2, 3, and 4 modes are plotted on the density maps of the $R$-$\phi$ plane. The $m=2$ mode traces the spiral arms better than the other modes.
Fig.~\ref{fig:hist_bar_angle_A2} shows the histograms for the $\phi$ of the positions with $\mathrm{KLD}<0.2$  for three spiral strength cases: $|A_2|<0.02$, $0.02<|A_2|<0.04$, and $0.04<|A_2|$. Here, the analysis is limited to the positions at $R=8\;\kpc$ and $8.5\;\kpc$. 
The shape of the histograms depends on the spiral strength. The histogram of $0.02<|A_2|<0.04$ has a peak at $\phi \simeq 30^{\circ}$ and a valley at $\phi \simeq 110^{\circ}$.  
The histogram of $|A_2|<0.02$ is less steep than the one of $0.02<|A_2|<0.04$. We see a plateau around $\phi=0^{\circ}$--$40^{\circ}$.
The histogram of $|A_2|>0.04$ is almost flat. When the spiral arm is strong, there is no specific angle where we often detect velocity distributions with small KLDs.

\begin{figure}
	\begin{center}
		\includegraphics[width=\columnwidth]{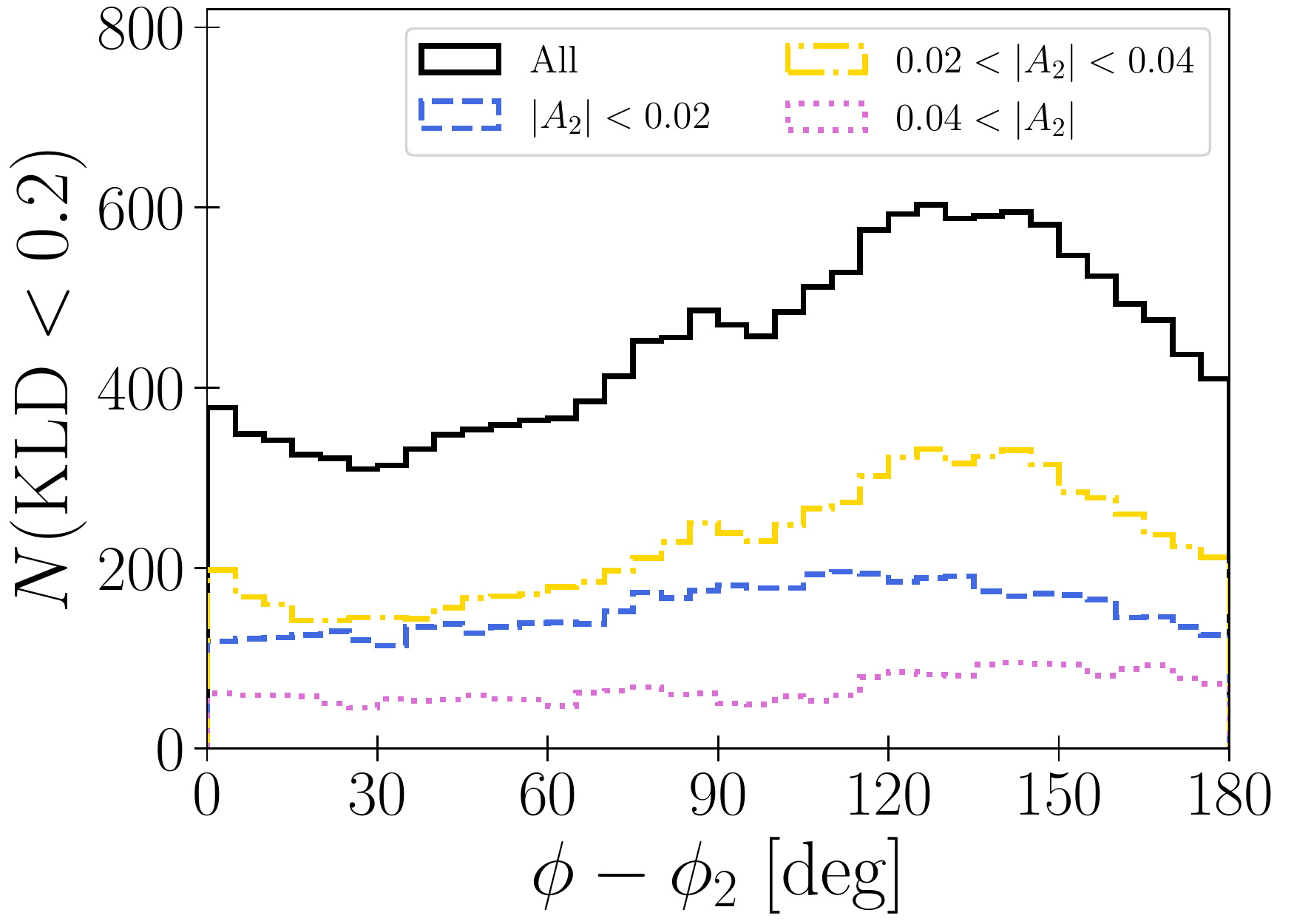}
		\caption{Same as Fig.~\ref{fig:hist_bar_angle_A2} but now for the angle with respect to the spirals ($\phi-\phi_2$).}\label{fig:hist_spiral_angle_A2}
	\end{center}
\end{figure}

Fig.~\ref{fig:hist_spiral_angle_A2} shows the same histograms as Fig.~\ref{fig:hist_bar_angle_A2} but now with respect to the spirals ($\phi-\phi_2$).
The histogram for all the spiral strengths (black line) shows a peak around $\phi=130^{\circ}$. The spiral positions are $\phi=0^{\circ}$ and $180^{\circ}$, therefore the peak is at the inter-arm region, which is consistent with observations. Very long baseline interferometry (VLBI) observations suggest that the Sun is located in the inter-arm regions of the MW's main spiral arms Perseus and Sagittarius-Carina \citep{2019ApJ...885..131R, 2020PASJ...72...50V}. 
We note that the Sun may be close to the `Local Arm', but that its features are not as clear as the Perseus or Sagittarius-Carina arms (see \citealt{2019ApJ...882...48M} and the references therein). 
The histograms in the cases of $|A_2|<0.02$ and $|A_2|>0.04$ are flatter than those of $0.02<|A_2|<0.04$.

	It is unclear why the velocity-space distributions with small KLDs are more frequently detected in the inter-arm regions than in the arm regions. One possible explanation is that the spiral arms disrupt the velocity-space substructure as formed by the bar resonances. However, spiral arms can also form velocity-space substructures as shown in  \citet{2021arXiv211115211K} where direct imprints of the spiral arms appear as velocity-space substructures. In Section~\ref{sec:discussions} we will see that the $N$-body model does not reproduce the detailed Hyades-Pleiades stream structures. Bar resonances cannot explain the origin of these structures and thus they may be due to the spiral arms.

\section{Discussions}\label{sec:discussions}
The previous section shows that the velocity-space distributions fluctuate with time in the simulation. However, some specific positions, namely $(R,\phi) \simeq (8\;\kpc, 20^{\circ})$, $(8.2\;\kpc, 30^{\circ})$, and $(8.5\;\kpc, 50^{\circ})$, frequently show velocity distributions similar to that in the solar neighbourhood.
The $(R,\phi)$ dependence on the KLD implies that the bar resonances influence the velocity-space distribution.
In this section, we discuss how the bar resonances impact the local velocity-space distributions at these positions.

\begin{figure*}
	\begin{center}
		\includegraphics[width=\columnwidth]{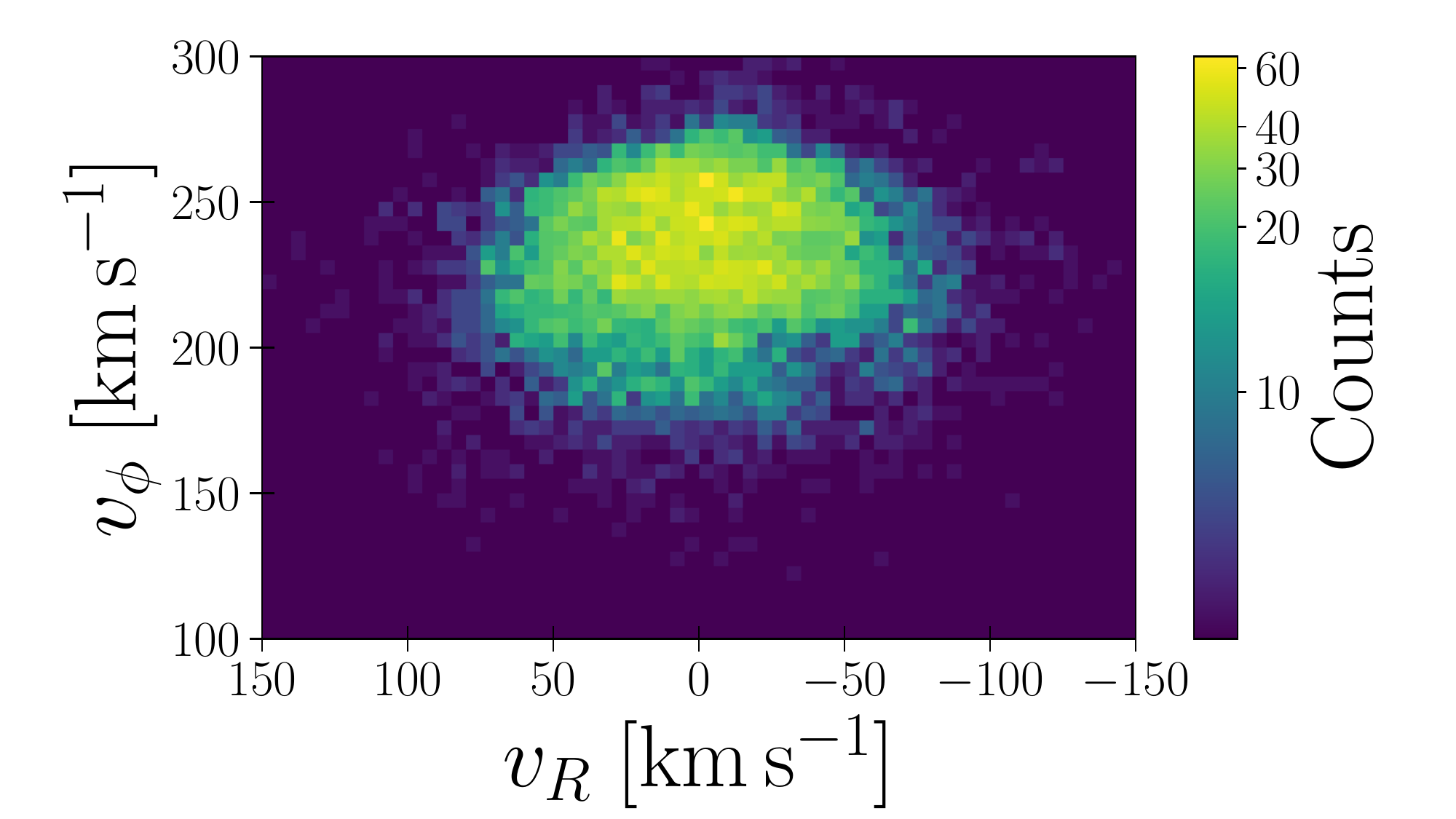}
		\includegraphics[width=\columnwidth]{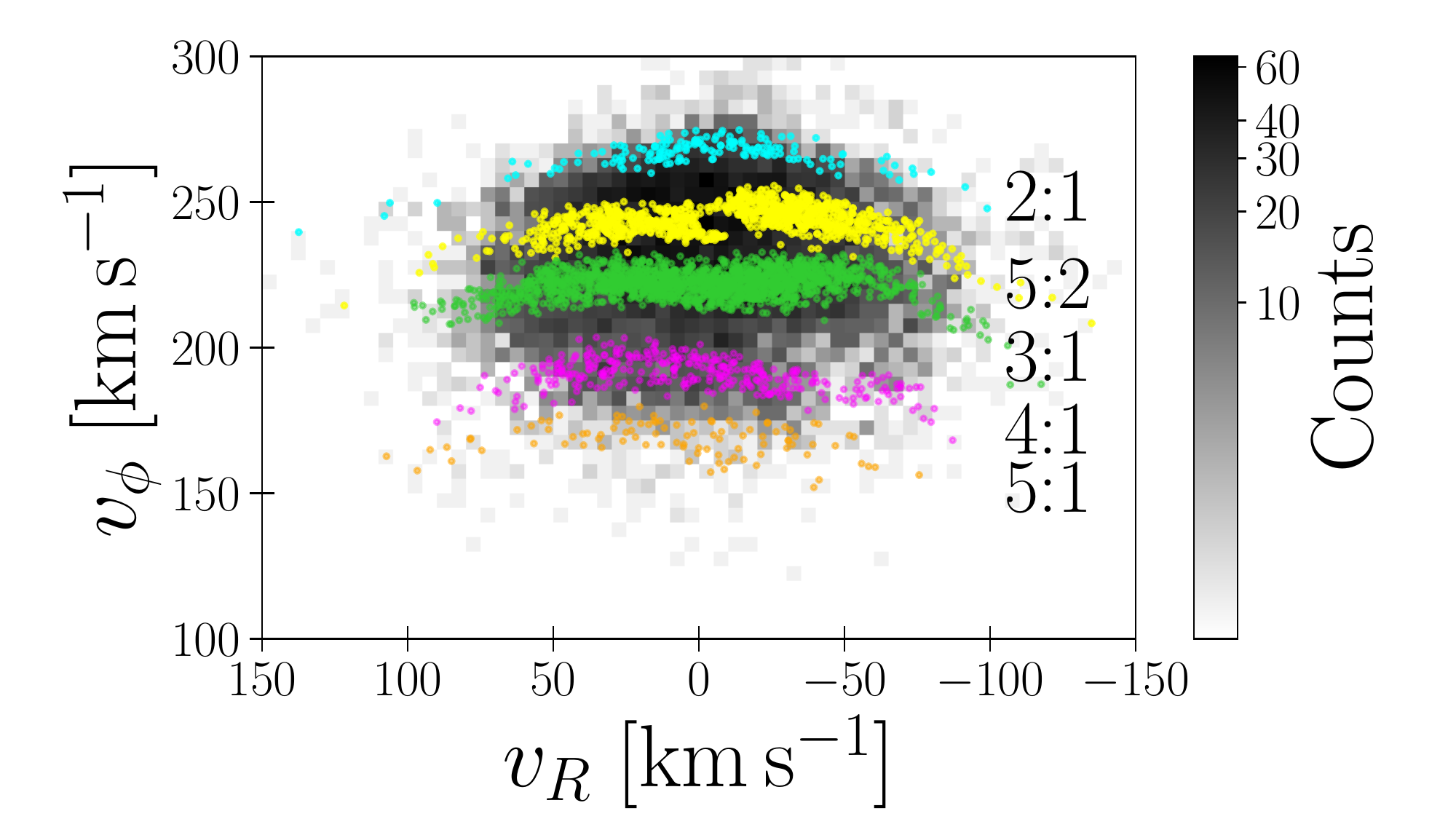}
		\caption{Velocity-space distribution for the particles within 200\;pc from $(R,\phi)=(8\;\kpc, 15^{\circ})$ at $t=9.29$\;Gyr. \textit{Left:} The colours indicate number counts in each bin. The bin size is $5 \times 5 \; {(\kms)}^2$. \textit{Right:} Velocity-space distributions of the particles trapped in bar resonances. Cyan, yellow, green, magenta, and orange dots indicate particles trapped in 2:1, 5:2, 3:1, 4:1, 5:1 OLRs, respectively.}\label{fig:uv_sim_8_15}
	\end{center}
\end{figure*}

Fig.~\ref{fig:uv_sim_8_15} shows the velocity-space space distributions for the particles within 200\;pc from $(R,\phi)=(8\;\kpc, 15^{\circ})$ at $t=9.29$\;Gyr. The KLD of this distribution is 0.145. This is one of the smallest values for the velocity-space distributions at $R=8\;\kpc$.
The map in the left panel of Fig.~\ref{fig:uv_sim_8_15} shows some substructures similar to that in Fig.~\ref{fig:uv_gaia}.
Hercules-like, horn-like, Sirius-like, and hat-like structures locate at $(v_R, v_{\phi}) \simeq (10,200)\;\kms$, $(-50,220)\;\kms$, $(-50,240)\;\kms$, and $(0,270)\;\kms$, respectively.
We compute the orbital frequencies of the particles around $(R,\phi)=(8\;\kpc, 15^{\circ})$ and select the ones trapped in bar resonances based on the frequency ratios. We mainly identify five resonances namely 2:1, 5:2, 3:1 ,4:1, and 5:1 OLRs. The right panel of Fig.~\ref{fig:uv_sim_8_15} shows their distributions in the velocity space. Cyan, yellow, green, magenta, and orange dots indicate particles trapped in 2:1, 5:2, 3:1, 4:1, 5:1 OLRs, respectively.
The trajectories of the resonant orbits are shown in Fig.~6 and the supplementary data of \citet{2020MNRAS.499.2416A}.
The Hercules-like stream is made from the 4:1 and 5:1 OLRs. Particles trapped in 2:1 and 3:1 OLRs contribute to the hat-like and horn-like structures respectively.
\citet{2020MNRAS.499.2416A} identified the same correspondence between the resonances and the velocity-space substructures around $(R,\phi) = (8\;\kpc, 20^{\circ})$  at $t=10$\;Gyr.
Although the 5:2 OLR is not prominent at $t=10$\;Gyr, there is still a large number of particles trapped in this resonance and they form a Sirius-like stream at $t=9.29$\;Gyr.

\begin{figure*}
	\begin{center}
		\includegraphics[width=\columnwidth]{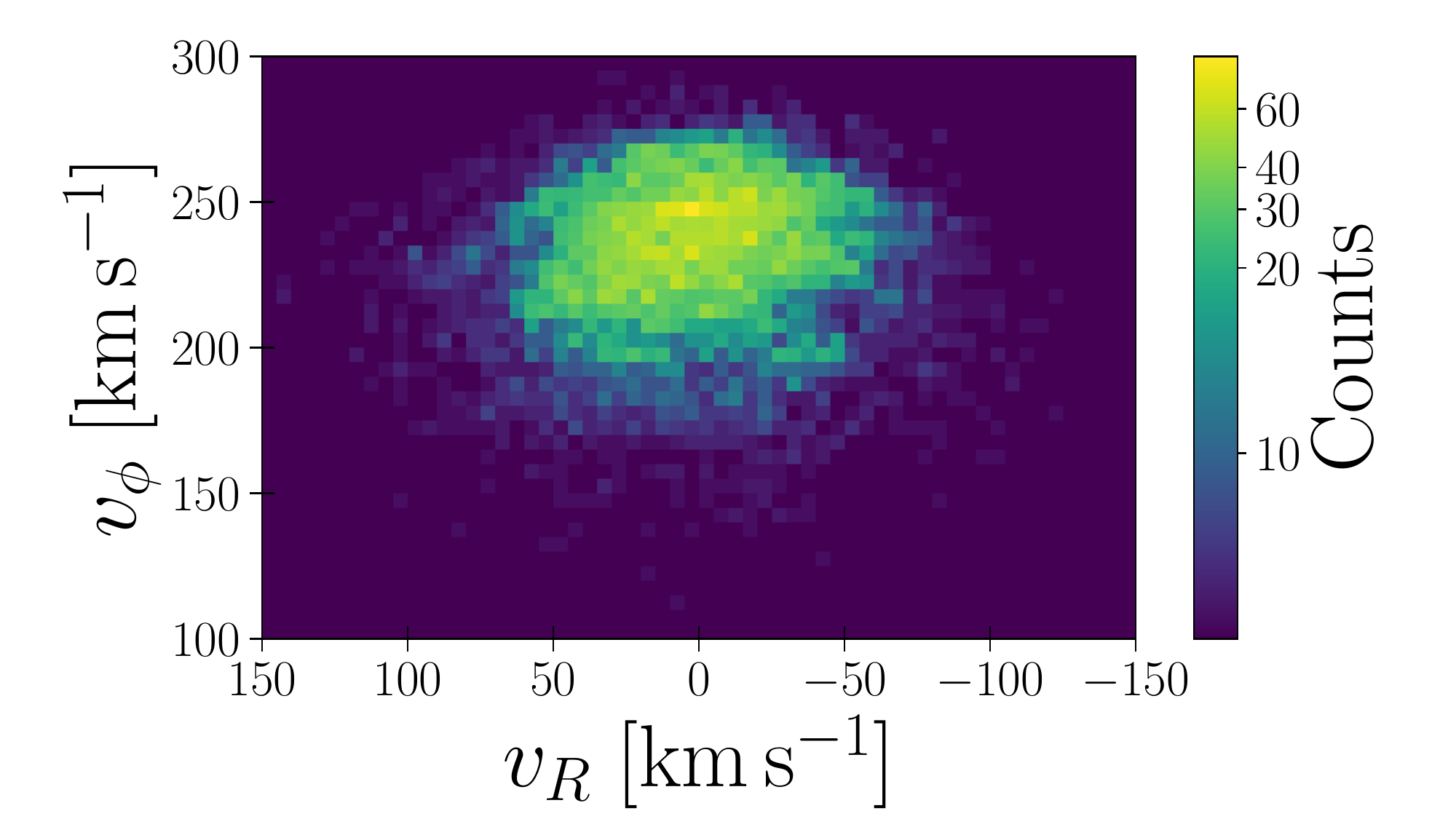}
		\includegraphics[width=\columnwidth]{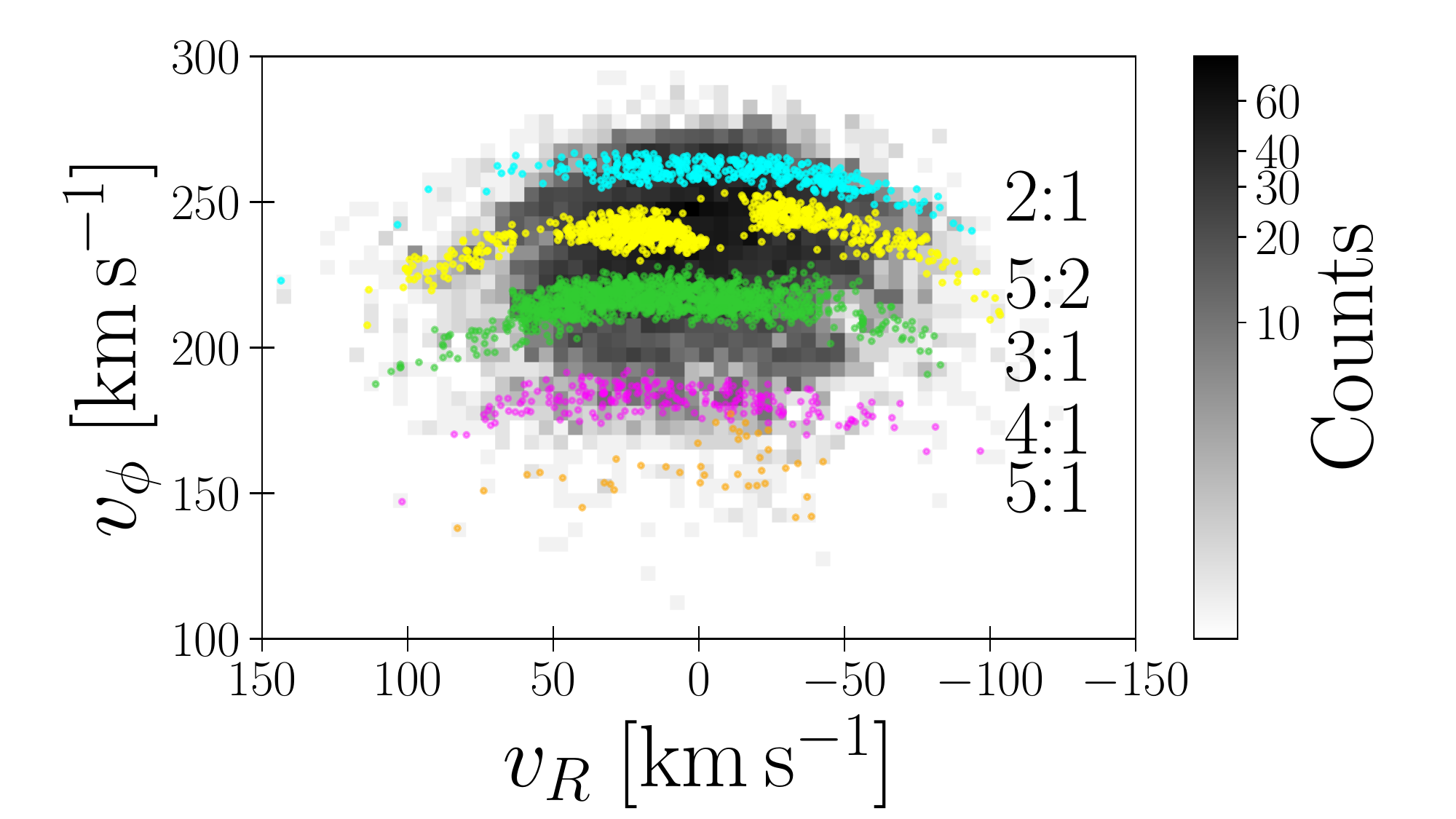}
		\caption{Same as Fig.~\ref{fig:uv_sim_8_15} but for the particles within 200\;pc from $(R,\phi)=(8.5\;\kpc, 50^{\circ})$ at $t=8.97$\;Gyr.}\label{fig:uv_sim_8p5_50}
	\end{center}
\end{figure*}

Fig.~\ref{fig:uv_sim_8p5_50} shows the same velocity-space distribution as Fig.~\ref{fig:uv_sim_8_15} but now for the particles within 200\;pc from $(R,\phi)=(8.5\;\kpc, 50^{\circ})$ at $t=8.97$\;Gyr. The velocity-space distribution also has one of the smallest KLDs of $\sim0.145$. Visible in the left panel are Hercules-like, horn-like, Sirius-like, and hat-like substructures.
In this case, the horn-like, Sirius-like, and hat-like substructures consist of particles trapped in 3:1, 5:2, and 2:1 OLRs, respectively. The right panel shows that the particles trapped in 4:1 and 5:1 OLRs are part of the Hercules-like stream, but the number of the particles in 5:1 OLR is smaller than that in Fig.~\ref{fig:uv_sim_8_15}. This is because $R=8.5$\;kpc is further from the 5:1 OLR radius. It is located at around $R\simeq6.5$\;kpc for the bar pattern speed of $\OmegaBar \simeq 45\;\kmskpc$, which is the typical value for the latter epochs in our simulation.

The resonances of odd modes such as 3:1, 5:1, and 5:2 OLRs are due to the asymmetry of the bar potential, which is a natural consequence of $N$-body simulations. Most of the studies using test particle simulations assume symmetric bar potentials, whose Fourier decompositions include only even modes. Bars in $N$-body models are not completely symmetric, and therefore odd-mode resonances arise.
	\citet{2019A&A...626A..41M}	studied the impact of higher-order bar resonances on the velocity-space distribution of stars. The method used and the details of the results are different from ours.
	In both theirs and our models the hat is made from 2:1 OLR. However, the correspondences between the other substructures and bar resonances are different from ours. In their model, Hercules, Serius, and horn, structures are made from CR, 4:1, and 6:1 OLR respectively. 
	In their model the Galactic potential is based on the M2M method \citep{2017MNRAS.465.1621P}. The bar's pattern speed is $\OmegaBar=39\;\kmskpc$, and the CR radius is located at $R\simeq 6\;\kpc$. On the other hand, the CR radius of our $N$-body model is $R\simeq 5\;\kpc$ and we do not observe particles in CR at $R\simeq8$--$8.5\;\kpc$.
	The bar potential of \citet{2019A&A...626A..41M} comprises the Fourier components of $m=2$, 3, 4, and 6. The lack of the 5:1/5:2 OLR might be due to the lack of the $m=5$ mode. 
	The rotation curves (i.e., the axisymmetric component of the Galactic potential) also affect the distribution of the resonantly trapped stars.

\begin{figure}
	\begin{center}
		\includegraphics[width=\columnwidth]{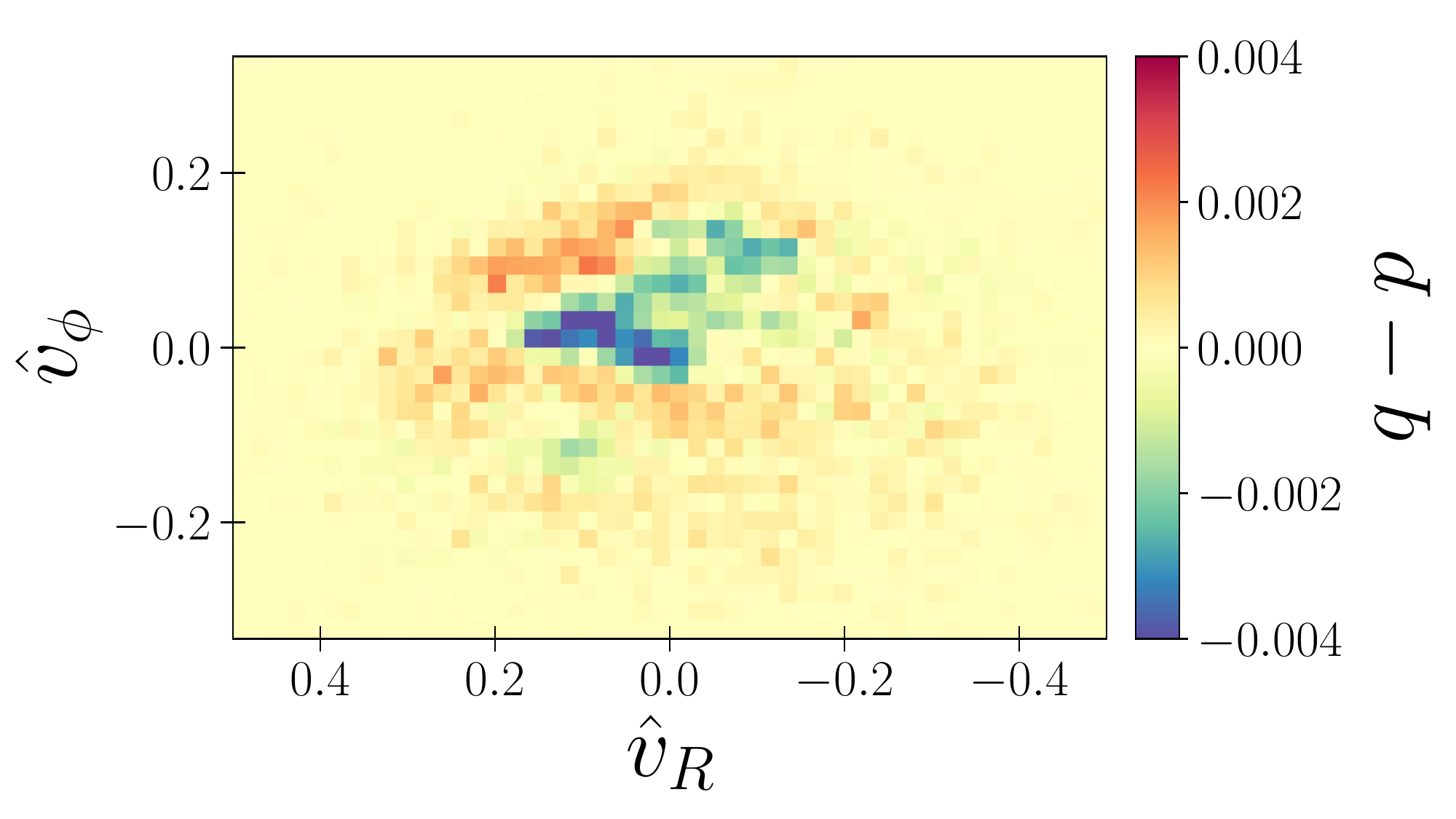}
		\includegraphics[width=\columnwidth]{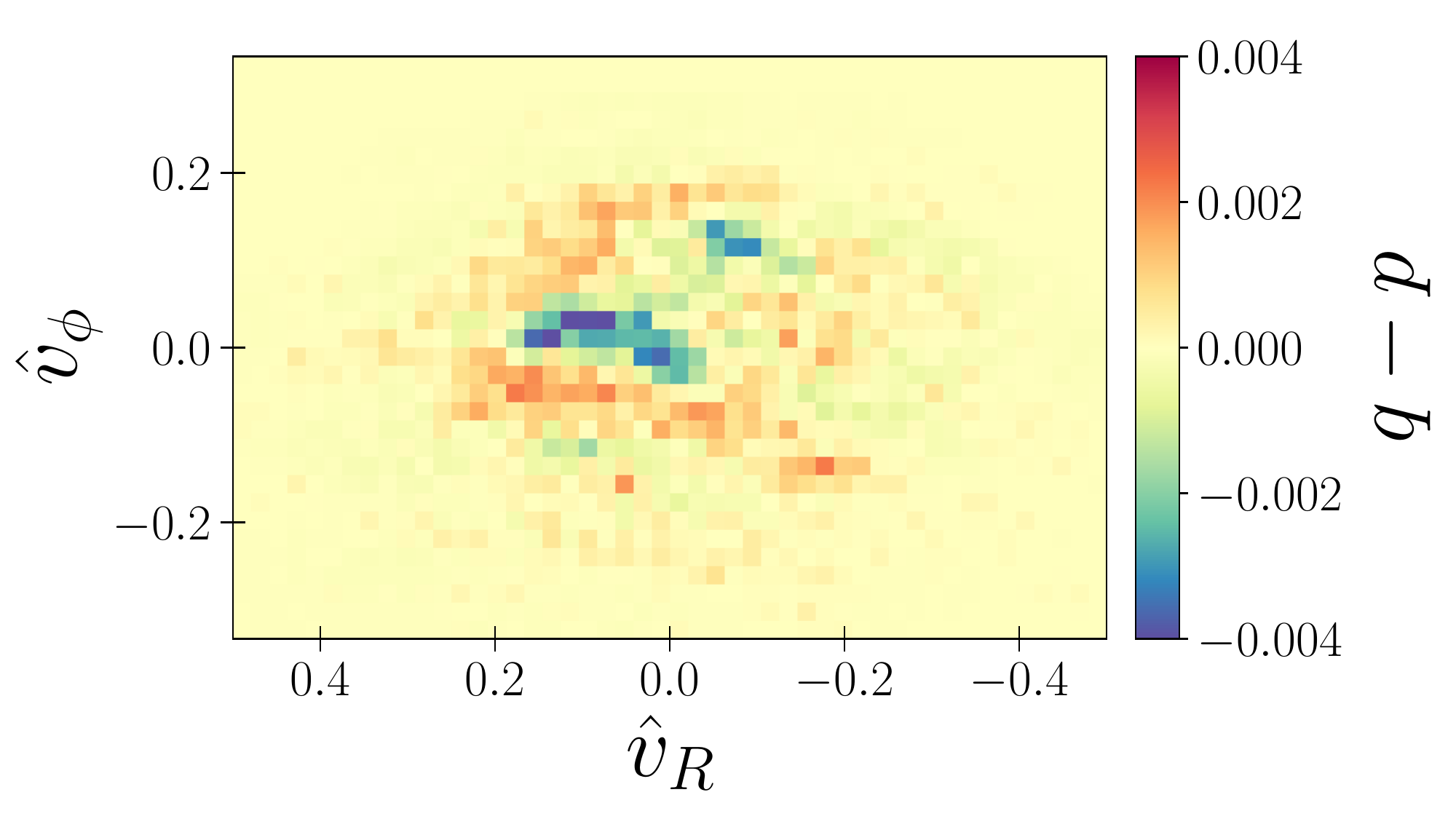}
		\caption{Residual maps of the simulated and observed $\hat{v}_R$-$\hat{v}_{\phi}$ space distributions.  $p$ and $q$ are the probability distributions in the observation and the simulation, respectively, as defined in Section~\ref{sSec:KLD_analysis}. \textit{Top:} $q$ is evaluated at $(R,\phi) = (8\;\kpc, 15\degr)$  at $t=9.29$\;Gyr. \textit{Bottom:} $q$ is evaluated at $(R,\phi) = (8.5\;\kpc, 50\degr)$  at $t=8.97$\;Gyr.}\label{fig:uv_residual}
	\end{center}
\end{figure}

Particles in the 3:1 OLR distribute on the Hyades-Pleiades region in addition to Horn. \citet{2022MNRAS.509..844T} also obtained a similar result that the location of 3:1 OLR's resonance line is close to the ridge of Hyades and Horn in action space. 
The compact two peaks corresponding to Hyades and Pleiades respectively are not clearly identified in the density maps of Fig.~\ref{fig:uv_sim_8_15} or Fig.~\ref{fig:uv_sim_8p5_50}. 
	The residual maps in Fig.~\ref{fig:uv_residual} highlight the differences of the velocity-space distributions between the observation and the simulation. The colours in the maps show the $q-p$ value at each of the points in the $\hat{v}_R$-$\hat{v}_{\phi}$ space. Here $p$ and $q$ are the probability distributions in the $\hat{v}_R$-$\hat{v}_{\phi}$ space derived from the \textit{Gaia} and simulation data respectively. The upper and lower panels are the residual maps for the velocity-space distributions at $(R,\phi)=(8\;\kpc, 15\degr)$ at $t=9.29$\;Gyr (i.e.,\, the left panel in Fig.~\ref{fig:uv_sim_8_15}) and $(R,\phi)=(8.5\;\kpc, 50\degr)$ at $t=8.97$\;Gyr (i.e.,\, the left panel in Fig.~\ref{fig:uv_sim_8p5_50}), respectively.
	We see the dense blue regions from $(\hat{v}_R, \hat{v}_{\phi}) \simeq(0.2, 0.1)$ to $(0, -0.1)$ in both maps. These are the velocity-space substructures that are not reproduced by the $N$-body model.
This may be due to the resolution limitation of the simulation. Another possibility is that they originate from mechanisms other than bar resonances such as spiral arms \citep{2005AJ....130..576Q, 2018ApJ...863L..37M, 2020ApJ...888...75B}.

Another difference between the velocity-space distribution in the observation and those in the simulation is the internal structure of the Hercules stream. \textit{Gaia} data shows a trimodal structure for the Hercules stream, which is not seen in the simulation.  The Hercules-like streams in our simulation consist of the two resonances of 4:1 and 5:1 OLRs but do not have a third component.

\section{Summary}\label{sec:summary}
In this paper, we have quantitatively measured the similarities of velocity-space distributions using the Kullback-Leibler divergence (KLD). We have evaluated the KLDs between the $v_R$-$v_{\phi}$ space distribution for the solar neighbourhood stars observed by the \textit{Gaia} and those in an $N$-body MW model simulated by \citet{2019MNRAS.482.1983F}. The KLDs in the simulation show time evolution and spatial variation.

First, we have evaluated the KLDs at the three fixed points of $(R,\phi)=(7\;\kpc, 25\degr)$, $(8\;\kpc, 25\degr)$, and $(9\;\kpc, 25\degr)$. The time evolution of the KLDs are linked with bar's evolution. The high KLDs (i.e.\ low similarities) at the beginning of the simulation reflect the initial condition. They drop rapidly around the bar formation epoch ($t\simeq3$\;Gyr). During the bar's slowing down phase ($3\;\mathrm{Gyr} \lesssim t \lesssim 7\;\mathrm{Gyr}$), they decrease with time. After the slowing down ($t\gtrsim 7$\;Gyr), the KLDs are almost constant but show small fluctuations. The small KLDs in this epoch indicate relatively high similarities. Especially, the KLD at $(R,\phi)=(8\;\kpc, 25\degr)$ is smaller than the other two positions. In this position of the simulation, we frequently but not always observe the velocity-space distributions similar to that of the solar neighbourhood.

Next, we have investigated where in the disc we often detect velocity-space distributions similar to that in the solar neighbourhood.
Velocity-space distributions with sufficiently high similarities ($\mathrm{KLD}<0.2$) are frequently found in the range of $R=8$--8.5\;kpc. The detection frequency at $R\lesssim7\;\kpc$ and $R\gtrsim9\;\kpc$ are almost zero.
The detection frequency depends also on $\phi$. When $R$ is fixed, there is a specific angle at which the small KLDs are detected most frequently. The peak angle moves in the direction of positive $\phi$ as $R$ increases. Especially at $R=8.2$\;kpc, the peak is $\phi\simeq30\degr$. This $R$ and $\phi$ are close to those of the Sun.
Spiral arms also impact the velocity-space distribution. The $(R,\phi)$ dependence of the KLD is weaken when the spiral arms are strong. 
Furthermore, the velocity-space distributions with small KLDs are more frequently detected at the inter-arm regions than the arms regions.

We have investigated the relation between the bar resonances and the substructures in the velocity distributions with small KLDs. We have plotted the resonantly trapped particles in the velocity map at $(R,\phi) =(8\;\kpc,15\degr)$ at $t=9.29$\;Gyr. 
We have performed the same analysis for the velocity map at $(R,\phi) =(8.5\;\kpc,50\degr)$ at $t=8.97$\;Gyr. In both the cases, Hercules-like, horn-like, Sirius-like, and hat-like substructure are confirmed. They are made from bar resonances. The Hercules-like streams consist of 4:1 OLR and 5:1 OLR.\@ Our previous study \citep{2020MNRAS.499.2416A} obtained the same conclusion from the analysis of the final snapshot only.
Bar's higher order resonances as origin of the phase-space substructures are discussed in other resent studies \citep[e.g.][]{2017MNRAS.471.4314M, 2019MNRAS.484.4540H, 2019A&A...626A..41M, 2021MNRAS.506.4687M,2022MNRAS.509..844T}.

As the KLD's oscillation suggests, the velocity-space distribution at a fixed position largely fluctuates. However, even in the non-static model, the bar resonances have significant impact on the stellar velocity-space distribution. Spiral arms may weaken the underlying influence of the bar resonances and cause the fluctuation of the KLD.\@ 
This is consistent with the result that the detection frequency of the small KLD is higher in the inter-arm regions than in the arm regions.

\section*{Acknowledgements}
We thank the anonymous referee for the useful comments.
We thank Kohei Hattori for useful discussions.
This work has made use of data from the European Space Agency (ESA) mission \textit{Gaia} (\url{https://www.cosmos.esa.int/gaia}), processed by the \textit{Gaia} Data Processing and Analysis Consortium (DPAC, \url{https://www.cosmos.esa.int/web/gaia/dpac/consortium}). 
Funding for the DPAC has been provided by national institutions, in particular the institutions participating in the \textit{Gaia} Multilateral Agreement.
Simulations are performed using GPU clusters, HA-PACS at the University of Tsukuba, Piz Daint at CSCS, Little Green Machine II (621.016.701) and the ALICE cluster at Leiden University.
Initial development has been done using the Titan computer Oak Ridge National Laboratory. This work was supported by a grant from the Swiss National Supercomputing Centre (CSCS) under project ID s548 and s716.
This research used resources of the Oak Ridge Leadership Computing Facility at the Oak Ridge National Laboratory, which is supported by the Office of Science of the U.S. Department of Energy under Contract No. DE-AC05-00OR22725 and by the European Union's Horizon 2020 research and innovation programme under grant agreement No 671564 (COMPAT project).

\section*{Data Availability}
The simulation data are available at \url{http://galaxies.astron.s.u-tokyo.ac.jp/}.



\bibliographystyle{mnras}
\bibliography{refs,mybib} 




\appendix
\section{Positions of spiral arms}\label{sec:spiral_positions}
\begin{figure}
	\begin{center}
		\includegraphics[width=\columnwidth]{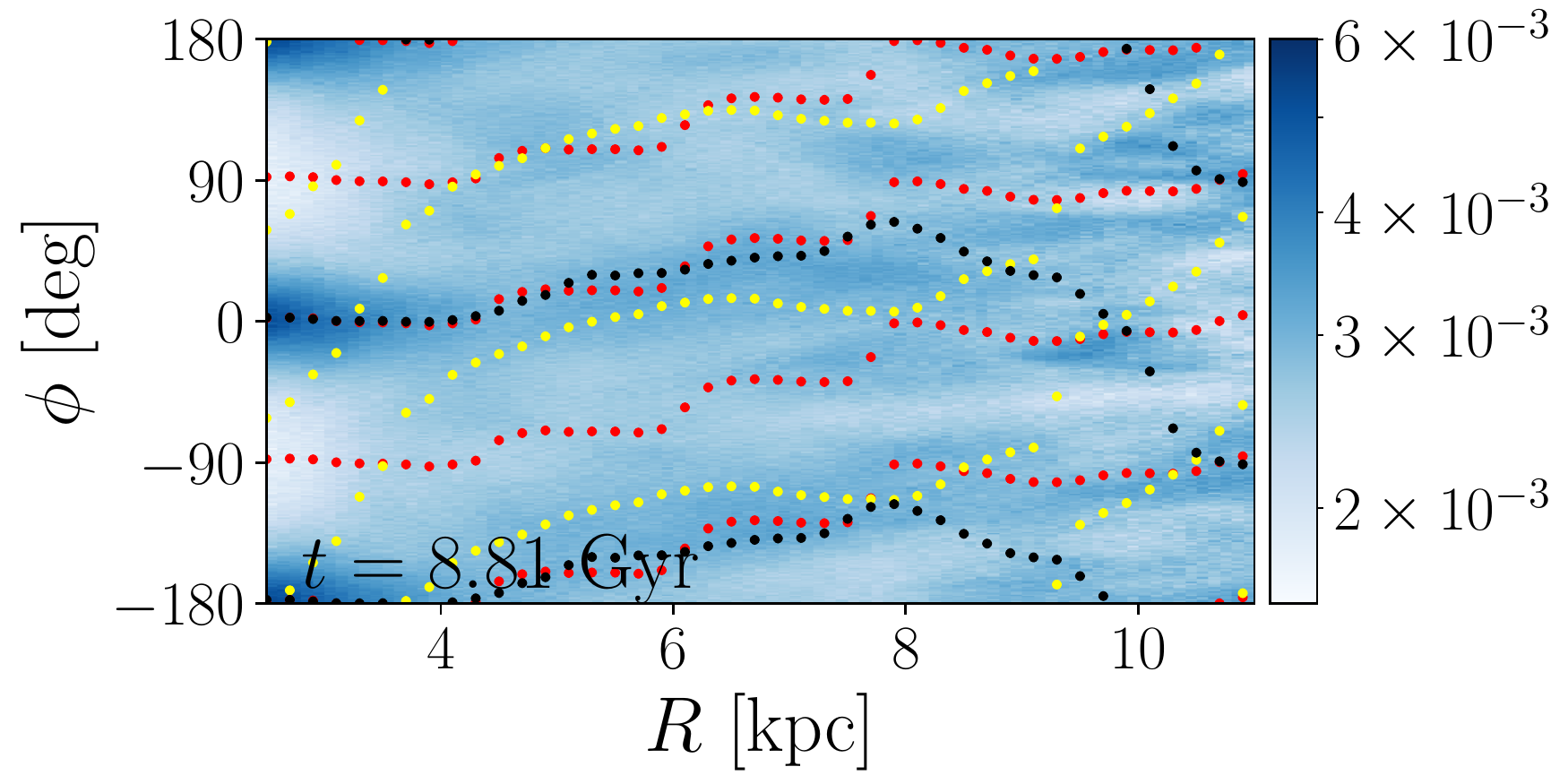}
		\includegraphics[width=\columnwidth]{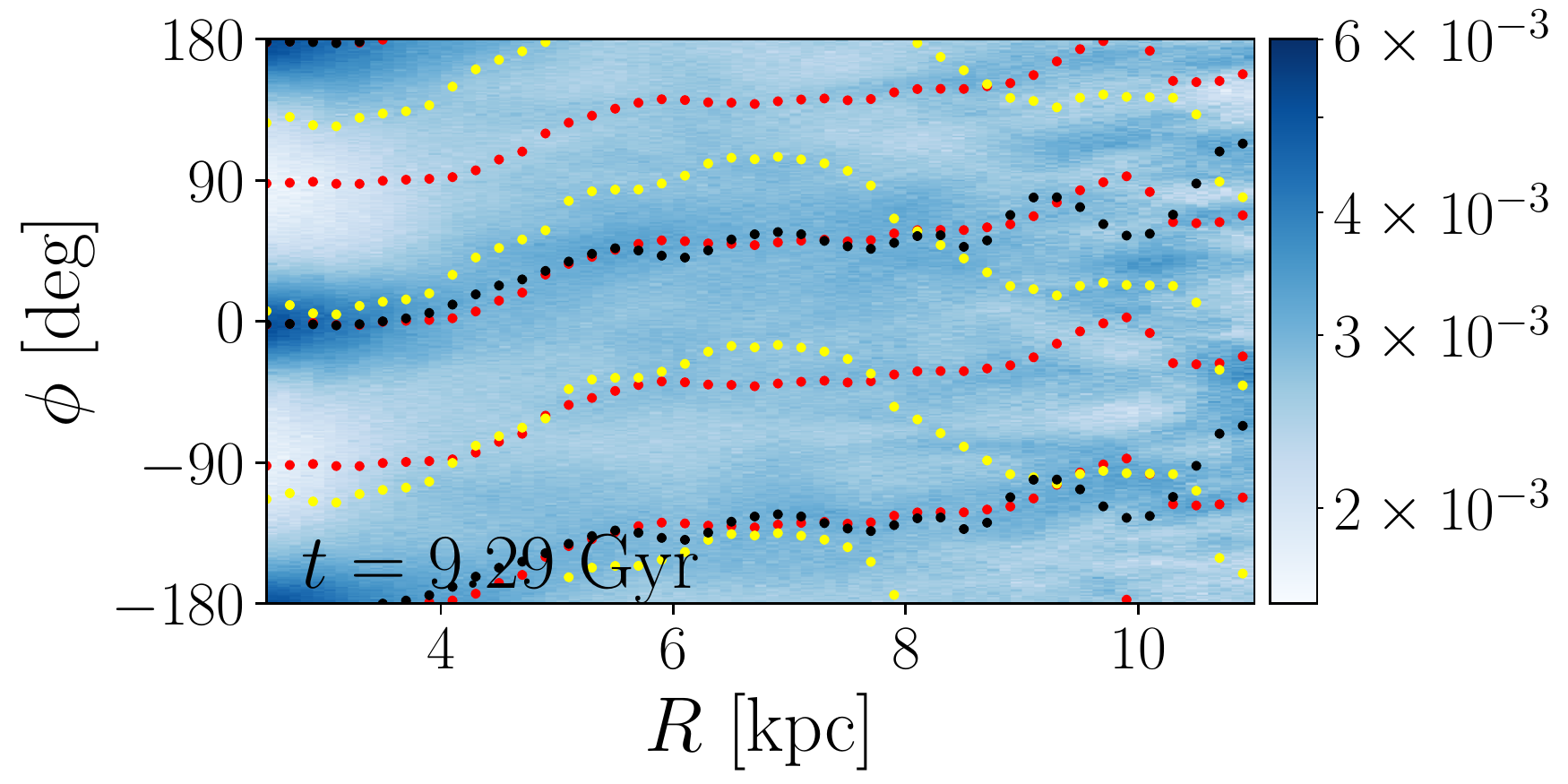}
		\includegraphics[width=\columnwidth]{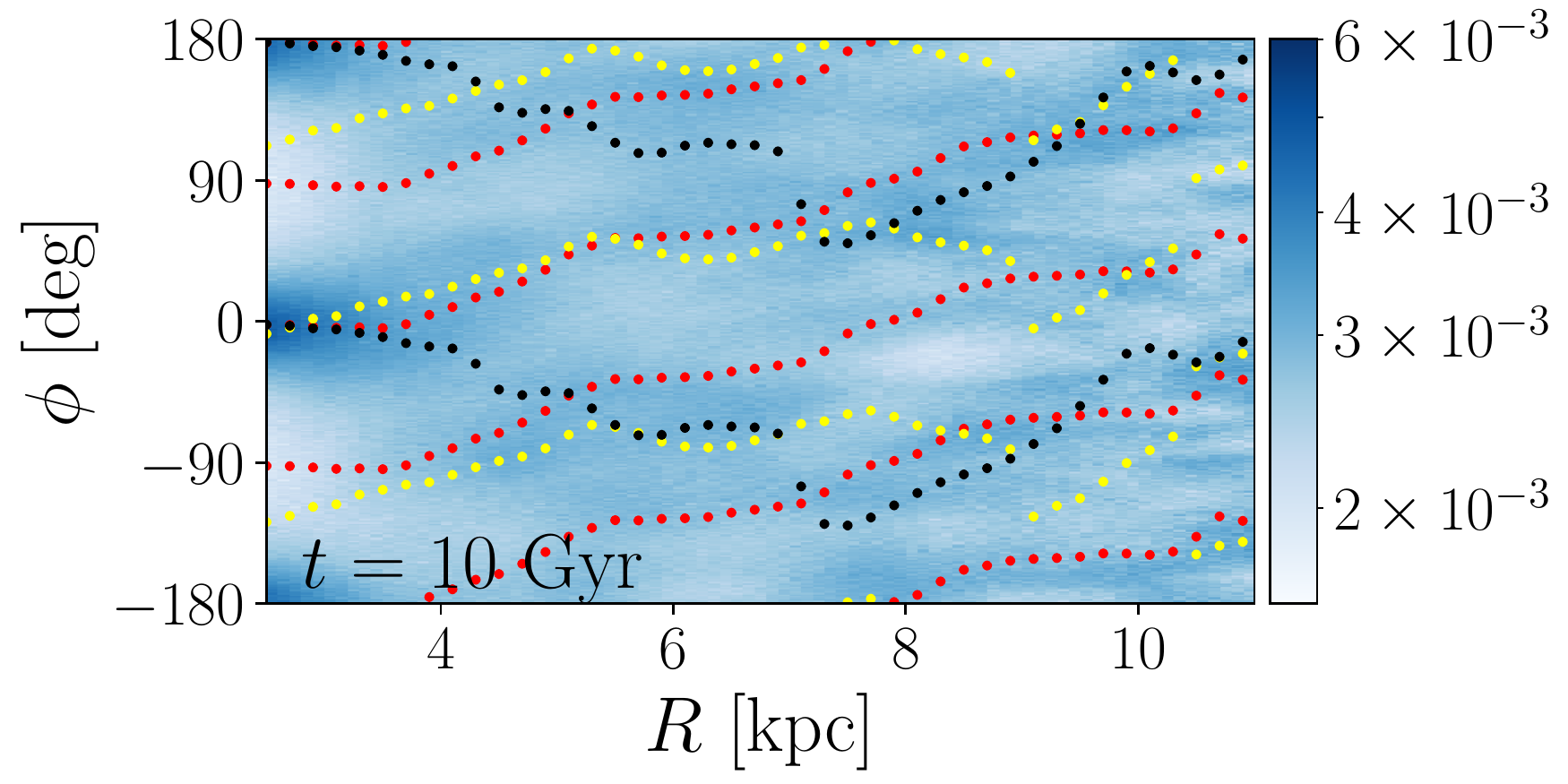}
	\end{center}
	\caption{The phase angles for the Fourier decomposition of the disc surface density, shown at $t=8.81$, 9,29, and 10\,Gyr. Black, yellow, and red dots represent the phase angles for the $m=2$, 3, and 4 modes, respectively. The background colour represent the normalized surface densities, $\Sigma(R,\phi)/\Sigma_0(R)$.}\label{fig:phase_angles}
\end{figure}

	In $N$-body simulations of disc galaxies the spiral are not in steady states. Instead they undergo repeated formation and destruction \citep{2013ApJ...763...46B}. and determining their positions is not straightforward. In this paper the Fourier decomposition of the disc surface density is used to determine their position. In Fig.~\ref{fig:phase_angles}, the phase angles $\phi_m(R)$ for $m=2$, 3, and 4 are overploted on the normalized density maps of the $R$-$\phi$ space. 
	The overdense regions (i.e., spiral arms) show complex morphologies. None of the Fourier modes completely traces the overdensities. However, the $\phi_2(R)$ fits the high-density regions relatively well and hence we use that to define the spiral arms positions.



\bsp
\label{lastpage}
\end{document}